\documentclass[reprint,10pt]{revtex4-1}

\usepackage{amsmath}    
\usepackage{bm}
\usepackage{graphicx}   
\usepackage{verbatim}   
\usepackage{color}      
\usepackage[hidelinks]{hyperref}   
\usepackage[caption=false]{subfig}
\usepackage{enumerate}

\mathchardef\mhyphen="2D

\begin{document}

\title{\bf{$R$-matrix with Time-dependence Theory for Ultrafast Atomic Processes in Arbitrary Light Fields}}

\author{D. D. A. Clarke}
\email{dclarke23@qub.ac.uk}
\author{G. S. J. Armstrong}
\author{A. C. Brown}
\author{H. W. van der Hart}
\affiliation{Centre for Theoretical Atomic, Molecular and Optical Physics, School of Mathematics and Physics, \\
 Queen's University Belfast, University Road, Belfast, BT7 1NN, Northern Ireland}

\date{\today}

\begin{abstract}

We describe an {\it ab initio} and non-perturbative $R$-matrix with time-dependence theory for ultrafast atomic processes in light fields of arbitrary polarization.\ The theory is applicable to complex, multielectron atoms and atomic ions subject to ultrashort (particularly few-femtosecond and attosecond) laser pulses with any given ellipticity, and generalizes previous time-dependent $R$-matrix techniques restricted to linearly polarized fields.\ We discuss both the fundamental equations, required to propagate the multielectron wavefunction in time, as well as the computational developments necessary for their efficient numerical solution.\ To verify the accuracy of our approach, we investigate the two-photon ionization of He, irradiated by a pair of time-delayed, circularly polarized, femtosecond laser pulses, and compare photoelectron momentum distributions, in the polarization plane, with those obtained from recent time-dependent close-coupling calculations.\ The predictive capabilities of our approach are further demonstrated through a study of single-photon detachment from F$^{-}$ in a circularly polarized, femtosecond laser pulse, where the relative contribution of the co- and counter-rotating $2p$ electrons is quantified.

\end{abstract}

\maketitle

\section{Introduction}

Precise control of the polarization state has become a key research directive for femtosecond and attosecond light-source technologies.\ In particular, in the extreme-ultraviolet (XUV) and soft-X-ray spectral ranges, intense and coherent elliptically polarized laser pulses have traditionally been realised only through large-scale facilities, such as femto-sliced synchrotrons \cite{Bahrtetal1992,Schoenleinetal2000,Cuticetal2011} and free-electron lasers \cite{Allariaetal2012,Allariaetal2014}.\ In recent years, however, a substantial effort has been made towards the development of compact alternatives, with the aim of meeting the requirements of photon-demanding applications, such as ultrafast metrology and spectroscopy, on a laboratory scale.\ In particular, solid-state and gas-based media, for high-order harmonic generation (HHG), continue to represent attractive means of producing ultrashort light pulses with manipulable polarization properties.\ To date, several schemes have been explored both experimentally, and theoretically, with fruition, relying on prealigned molecules as targets \cite{Levesqueetal2007,Zhouetal2009,Mairesseetal2010}, bichromatic and co- or counter-rotating drivers \cite{Milosevicetal2000,Fleischeretal2014,Medisauskasetal2015,Kfiretal2015}, cross-polarized, multicolor laser light \cite{Eichmannetal1995,Ruizetal2009,Lambertetal2015}, HHG assisted with static electric fields \cite{Borcaetal2000,YuanBandrauk2011}, and even resonance effects inherent in the dynamics of HHG itself \cite{Ferreetal2014}.\ Moreover, advances in plasma-based laser technology have enabled the tabletop demonstration of stable, circularly polarized, soft-X-ray pulses, with photon fluxes superior to current HHG sources \cite{Depresseuxetal2015}.

The development, and increasingly widespread availability, of polarization-tuneable light sources has facilitated a host of experimental opportunities, both for the investigation, and novel control, of laser-matter interactions on an ultrafast timescale.\ Indeed, the use of elliptically polarized fields opens the possibility of probing dynamical effects, and target properties, that may be inaccessible with linearly polarized pulses alone.\ In particular, recent experiments on atomic and molecular strong-field ionization, effected with circularly and/or elliptically polarized, femtosecond pulses, have revealed a number of new and exciting phenomena.\ Several authors \cite{Eckleetal2008Science,Martinyetal2009,Xieetal2017} have reported ``counterintuitive'' shifts in photoelectron angular distributions (attributable to a dynamical phase shift \cite{Proninetal2009}), as well as imprints of target orbital characteristics therein \cite{Martinyetal2010,Hansenetal2011,Dimitrovskietal2011,Abusamhaetal2011}.\ Coherent, circularly polarized laser pulses have also led to the observation of kinematic vortex patterns in momentum spectra \cite{Pengeletal2017}, and to the detection of spin-polarized electrons, created by non-adiabatic tunnelling \cite{Herathetal2012,Hartungetal2016}.\ Moreover, the potential of such pulses to serve as an attoclock, for timing attosecond-scale ionization dynamics, has been demonstrated via the angular streaking technique \cite{Eckleetal2008}, yielding unprecendented insight into the nature of quantum tunnelling \cite{Eckleetal2008Science,Staudteetal2009,Pfeifferetal2012,Wuetal2012,Torlinaetal2015}.

In order to complement these experimental efforts, theoretical treatments of the laser-atom interaction are compelled to address new challenges.\ Crucially, laser fields with non-zero ellipticity drive atomic dynamics that is intrinsically multidimensional in nature, largely invalidating simplified, reduced-dimensionality models of strong-field ionization \cite{Eberlyetal1988,Ruizetal2006}.\ More fundamentally, the conservation of the total orbital magnetic quantum number, $M_{L}$, that prevails for linear polarization is now lost:\ even in the dipole approximation, elliptically polarized fields effect transitions in which $M_{L}$ must change by $\pm 1$.\ Accounting for all such transitions (sometimes referred to as the $M_{L}$-mixing problem in the literature \cite{Muller1997,Kjeldsenetal2007}) can render first-principles, quantum dynamic simulations computationally intensive.

As a consequence of this complexity, only a limited number of {\it ab initio} approaches have been developed for atomic systems in arbitrarily polarized light fields.\ First-principles calculations have been reported for the H atom exposed to circularly polarized, femtosecond pulses \cite{Gajdaetal1994,Huensetal1997,Askelandetal2011,Baueretal2014}, where the time-dependent Schr\"{o}dinger equation (TDSE) was solved numerically by means of spectral methods.\ More recently, a computational technique based on the time-dependent close-coupling (TDCC) formalism \cite{Pindzolaetal2007,ColganPindzola2012}, combined with Wigner frame transformations \cite{Muller1997,Kjeldsenetal2007}, was applied to the laser-driven, two-electron problem, enabling the six-dimensional TDSE to be solved for a He atom subject to elliptically polarized, attosecond pulses \cite{Djiokapetal2015,Djiokapetal2016}.\ These approaches have offered some of the most detailed insight into the strong-field dynamics of one- and two-electron systems, revealing a differential response of co- and counter-rotating electrons \cite{Baueretal2014}, unusual manifestations of Ramsey interference \cite{Djiokapetal2015,Djiokapetal2016}, as well as a nonlinear dichroic effect in double-ionization \cite{Djiokapetal2014}.

Although full-dimensionality treatments have been effective in describing both single- and double-ionization phenomena, driven by intense, elliptically polarized laser light, their application to systems with more than two electrons is computationally intractable.\ Thus far, time-dependent simulations for multielectron atoms, exposed to ultrashort pulses with arbitrary polarization, have been rooted in the single-active-electron (SAE) approximation \cite{Martinyetal2009,Martinyetal2010,Abusamhaetal2011,Keldysh1965,Arboetal2008,Bauer2012,Popruzhenko2014}.\ Such approaches have played a key supporting role for experimental efforts, particularly where the photoelectron emission characteristics have been of primary interest.

Whilst the value of SAE methods cannot be denied, it has long been established that the dynamics of complex atoms, exposed to ultrashort laser pulses, are fundamentally many-body in nature \cite{Weberetal2000,Drescheretal2002,Uiberackeretal2007,FariaLiu2011,Leoneetal2014}.\ Recently, experiments have begun to uncover how spatially correlated electronic motion, previously probed with somewhat limited pulse-polarization control, might be manifest, and perhaps even coherently manipulated, with multidimensional light-fields \cite{Eckleetal2008Science,Torlinaetal2015,Shafiretal2012,Pfeifferetal2012,Kangetal2018}.\ It should be emphasized that adequate modelling approaches for such experiments are largely lacking:\ SAE (one-body) techniques can offer only a limited insight, and a computationally feasible, yet truly multielectron (many-body) treatment, has yet to be realized.

If theory is to play a complementary role for state-of-the-art experiments in strong-field physics, then sophisticated methods of calculation are required, offering an accurate account of both multielectron correlations in atomic structure, and a multielectron response to the light field.\ To this end, $R$-matrix with time-dependence (RMT) theory \cite{Mooreetal2011} offers an {\it ab initio} and non-perturbative technique for solving the TDSE, appropriate to general, multielectron atoms and atomic ions in strong laser fields.\ It represents the latest embodiment of a time-dependent R-matrix formalism \cite{BurkeBurke1997,Guanetal2007,Lysaghtetal2009}, whose flexibility and generality have been reflected in a wide variety of recent applications.\ These include multielectron correlation in doubly and core-excited states of Ne \cite{Dingetal2016}, strong-field rescattering in F$^{-}$ \cite{Hassounehetal2015} and HHG from noble gas atoms in the near-infrared regime \cite{Hassounehetal2014}.\ RMT theory has even been extended for the description of double-electron continua \cite{Wraggetal2015,WraggvdH2016}.\

In this article, we present a recent evolution in the RMT methodology, facilitating the analysis of ultrafast atomic dynamics in entirely arbitrary light fields.\ This capability subsumes the previous RMT methodology that was tailored specifically for linearly polarized laser pulses \cite{Mooreetal2011}.\ The generalization has been achieved by relaxing the constraint of $M_{L}$ conservation, allowing transitions, among different $LS$-coupled states of the target, in which $\Delta M_{L} = 0, \pm 1$.\ Not only does this accomodate arbitrary orientations for the axis of linear polarization, but also enables the adoption of laser pulses with circular or, more generally, elliptical polarization in RMT computations.\ We detail our extension of the existing RMT formalism, and associated computer codes, in Section {\ref{sec:Theory}}.\ As a means of verifying the accuracy of our new approach, we investigate, in Section {\ref{sec:He}}, the formation of multistart, spiral vortex features in the photoelectron momentum distributions of He, irradiated by a pair of time-delayed, ultrashort, circularly polarized laser pulses with opposite helicities.\ We compare our predicted photoelectron momentum distributions, in the polarization plane, with those obtained in a recent time-dependent close-coupling study.\ The predictive capabilities of our generalized RMT method are further demonstrated, in Section {\ref{sec:Fminus}}, through a study of single-photon detachment from F$^{-}$ in a circularly polarized, femtosecond laser pulse, where we quantify the sensitivity of the dynamics to the sign of the bound-electron magnetic quantum number.\ Section {\ref{sec:Conclusions}} closes the article with relevant conclusions.\ Finally, we note that atomic units are assumed throughout this work, unless otherwise stated.

\section{Theoretical and Computational Approach}
\label{sec:Theory}

We consider an atomic system, possessing $N+1$ electrons and nuclear charge $Z$, interacting with an intense and ultrashort light pulse of arbitrary polarization.

\subsection{Laser Field}

Throughout, the laser field is treated classically, and the interaction with the atomic system described in the dipole approximation.\ The electric field of a single, arbitrarily polarized laser pulse may be expressed in the form
\begin{equation}\label{field}
{\bf E}(t) = F(t)\textrm{Re}[ {\bf e}\,e^{-i(\omega t + \varphi)}],
\end{equation}
where $F(t)$ specifies the temporal envelope, $\omega$ is the carrier frequency, $\varphi$ is the carrier-envelope phase (CEP), and ${\bf e}$ is the polarization vector.\ In general, the vector ${\bf e}$ is complex $({\bf e}^{*}\cdot {\bf e} = 1)$, and can be written in the form
\begin{equation}
{\bf e} = ( \hat{ {\pmb{\epsilon}} } + i\eta\hat{ {\pmb{\zeta}} })/\sqrt{ 1 + \eta^{2} }, \nonumber
\end{equation}
where $-1 \leq \eta \leq 1$, $\hat{ {\pmb{\zeta}} } = \hat{ \bf k } \times \hat{ {\pmb{\epsilon}} }$, and $\hat{ \bf k }$, $\hat{ {\pmb{\epsilon}} }$ and $\hat{ {\pmb{\zeta}} }$ indicate, respectively, the propagation direction of the pulse and the major and minor axes of the polarization ellipse.\ In particular, for a linearly polarized field, $\eta = 0$, whilst for a right-hand (left-hand) circularly polarized field, $\eta = 1$ ($\eta = -1$).\ For a general, elliptically polarized pulse, $\vert \eta \vert$ serves to quantify the ellipticity.\

For calculations incorporating elliptically (and especially circularly) polarized laser fields, we shall often adopt $ \hat{ {\pmb{\epsilon}} } = \hat{{\bf x}}$, $\hat{ {\pmb{\zeta}} } = \hat{{\bf y}}$ and $\hat{{\bf k}} = \hat{{\bf z}}$ (see, for instance, Sections {\ref{sec:He}} and {\ref{sec:Fminus}}).\ However, the formalism discussed here is entirely general, and applies for any choice of propagation direction and orientation of the polarization plane.\

\subsection{TDSE}

Neglecting relativistic effects, the behaviour of the atomic system in the presence of the laser field, described by the time-dependent and multielectron wavefunction $\Psi({\bf X}_{N+1},t)$, is governed by the TDSE,
\begin{equation}\label{TDSE}
i\frac{\partial}{\partial t}\Psi({\bf X}_{N+1},t) = [H_{N+1} + D_{N+1}(t)]\Psi({\bf X}_{N+1},t).
\end{equation}
Here, $H_{N+1}$ is the field-free Hamiltonian,
\begin{equation}\label{FFHam}
H_{N+1}=\sum_{i=1}^{N+1}\left(-\frac{1}{2}\nabla_{i}^{2}-\frac{Z}{r_{i}}+\sum_{i>j=1}^{N+1}\frac{1}{r_{ij}}\right),
\end{equation}
and $D_{N+1}(t)$ is the dipole interaction operator, for $N+1$ electrons, in the length gauge,
\begin{equation}
D_{N+1}(t) = {\bf E}(t) \cdot \sum_{i=1}^{N+1} {\bf r}_{i}. \nonumber
\end{equation}
In these equations, we have regarded the target nucleus (assumed of infinite mass) to be located at the origin of coordinates, and we have written $r_{ij} = \vert {\bf r}_{i} - {\bf r}_{j} \vert$, where ${\bf r}_{i}$ and ${\bf r}_{j}$ are the position vectors of electrons $i$ and $j$.\ Also, ${\bf X}_{N+1} = {\bf x}_{1}, {\bf x}_{2},..., {\bf x}_{N+1}$, where ${\bf x}_{i} = {\bf r}_{i}\sigma_{i}$ denotes collectively the space and spin coordinates of electron $i$.\ We highlight that adoption of the length gauge, in the present theory, stands in contrast to previous strong-field calculations for one- and two-electron systems, wherein the velocity gauge was deemed advantageous \cite{Cormieretal1996}.\ However, on the basis of earlier investigations, conducted with time-dependent $R$-matrix theory \cite{Hutchinsonetal2010}, we have found that the interaction of a laser field with a multielectron atom is most accurately described in the length gauge.\ The velocity form of the dipole interaction operator appears to be less appropriate for such systems, since it emphasises the behaviour of the multielectron wavefunction at short-range.\ As a result, the latter requires a much superior description of atomic structure relative to the length form, and thus places a greater computational demand on time-dependent simulations of atomic strong-field processes.\

In extending the capabilities of the RMT approach, we have nonetheless preserved the essential philosophy of the method \cite{Mooreetal2011}.\ We therefore provide only a brief summary of the basic principles here, and devote Sections \ref{subsec:Inner} and \ref{subsec:Outer} to a discussion of the key developments, and modifications, of the original formulation.\ To enable a computationally efficient solution of (\ref{TDSE}), we employ the traditional $R$-matrix paradigm of dividing configuration space into two separate regions.\ This partition is effected with respect to the radial coordinate of an ejected electron, and yields an inner region, containing the target nucleus, and an outer region of relatively large radial extent.\ Within the inner region, multielectron exchange and correlation effects are accounted for in the construction of the many-body wavefunction.\ In the outer region, the ionised electron is regarded as sufficiently distant from the residual ion that exchange may be neglected.\ This electron is thus subject only to the long-range, multipole potential of the residual system, as well as the applied laser field.\ Importantly, RMT relies on a hybrid numerical scheme, consisting of a unique integration of basis set and finite-difference techniques.\ This enables the most appropriate method for solving the TDSE to be applied in each region.\

Whereas previous implementations of time-dependent R-matrix theory relied on a low-order Crank-Nicolson propagator, together with the solution of a system of linear algebraic equations \cite{Lysaghtetal2009}, the RMT approach adopts a high-order Arnoldi scheme \cite{Smythetal1998}.\ This replaces the solution of a linear system with a series of matrix-vector multiplications, which may reduce the numerical error in both the temporal and spatial propagation of the wavefunction.\ Since the Arnoldi algorithm is dominated by such operations, the RMT methodology offers substantially improved parallel scalability, making feasible calculations that exploit massively parallel computing resources (with more than 10000 cores).\

\subsection{Inner Region}
\label{subsec:Inner}

To solve Eq.\ (\ref{TDSE}) in the inner region, we expand the time-dependent, $(N + 1)$-electron wavefunction in a basis comprising eigenfunctions, $\psi_{k}({\bf X}_{N+1})$, of the field-free Hamiltonian ({\ref{FFHam}),
\begin{equation}\label{basis}
\Psi({\bf X}_{N+1},t) = \sum_{k} C_{k}(t)\psi_{k}({\bf X}_{N+1}).
\end{equation}
Note that the time-dependence is incorporated purely in the coefficients $C_{k}(t)$, such that they alone characterize the temporal evolution of the multielectron wavefunction.\ The basis functions $\psi_{k}({\bf X}_{N+1})$ are, in turn, developed in a close-coupling with pseudostates expansion \cite{BurkeBerrington1993,Burke2011}, generated from the $N$-electron wavefunctions of the residual ion states, as well as from a complete set of one-electron continuum functions, describing the motion of the ejected electron.\ Additional $(N + 1)$-electron correlation functions can be added to improve the quality of the basis set.\

As in the original formulation of RMT theory, it can be shown \cite{Mooreetal2011} that the time-dependent coefficients satisfy a system of first-order, ordinary differential equations,
\begin{equation}\label{ODEs}
\frac{d}{dt} C_{k}(t) = -i\sum_{k'}H_{kk'}(t)C_{k'}(t) + \frac{i}{2}\sum_{p} \omega_{pk}\left. \frac{\partial}{\partial r}f_{p}(r,t)\right\rvert_{r = b}.
\end{equation}
The quantities $H_{kk'}(t)$ are the matrix elements of the inner-region Hamiltonian, computed with respect to a basis consisting of the functions $\psi_{k}({\bf X}_{N+1})$.\ Furthermore, $\omega_{pk}$ are surface amplitudes, defined in \cite{Mooreetal2011}, and the functions $f_{p}(r,t)$ are the reduced radial wavefunctions of the ejected electron, found by resolution of the outer-region problem (see Section \ref{subsec:Outer}).\ The inner-region boundary radius is chosen as $r = b$.\ Note that the inhomogeneous nature of Eq.\ (\ref{ODEs}) arises due to inclusion of a Bloch operator in the analysis \cite{BurkeBerrington1993,Burke2011}, which suitably enforces hermiticity of the inner-region Hamiltonian.\ In fact, the second term on the right-hand side of this equation plays a critical role in RMT theory, for it connects a multielectron wavefunction in the inner region with a wavefunction that, at the boundary, is one-electron in nature and which, numerically, is obtained from the outer region.\

Thus far, our formulation of the generalized RMT theory follows that for purely linearly polarized laser light \cite{Mooreetal2011}.\ However, the complexities of modelling atomic systems, subject to multidimensional light fields, are already inherent in Eq.\ (\ref{ODEs}).\ As mentioned previously, the RMT approach relies on an efficient, high-order Arnoldi scheme to solve this system of equations, and thereby propagate the inner-region wavefunction in real-time.\ The latter entails numerical evaluation of a series of matrix-vector products, involving powers of the Hamiltonian matrix $(H_{kk'})$.\ Crucially, the precise structure of this matrix depends on the polarization state of the radiation, reflecting the dipole selection rules that dictate the admissible atomic transitions.\ Thus, to enable the treatment of truly arbitrary light-field configurations within the RMT framework, we must devise a single, robust computational strategy, facilitating an accurate solution of Eq.\ (\ref{ODEs}) for any relevant set of dipole selection rules.

We elaborate on this latter observation, and its implications for the generalization of the RMT approach, through a specific example, pertaining to a neutral, noble gas atom in two different light-field configurations.\ In particular, we compare the structure of the matrix $(H_{kk'})$ for a linearly polarized (one-dimensional) field, a case for which RMT had originally been formulated, and an arbitrarily polarized (three-dimensional) field.\ This comparison serves not only to highlight the essential modifications for the inner-region computations, but also to emphasize those features of the RMT method which render it most appropriate (over previous $R$-matrix techniques) for the advancements that we report here.\ Whilst, in the following discussion, we confine attention to the noble gas systems, it should be emphasized that these are merely exemplary.\ Indeed, the present methodology is applicable to entirely general multielectron atoms and atomic ions, offering the same flexibility, with respect to the choice of target, as its predecessors \cite{Lysaghtetal2009,Mooreetal2011}.

{\it (a) Linearly polarized laser field:} Firstly, we consider a linearly polarized field, whose axis is aligned along that of angular momentum quantization (typically the $z$-axis).\ In this case, the only permissible radiative transitions are those such that the change in the total orbital angular momentum quantum number $L$, and that in the quantum number associated with its projection $M_{L}$, satisfy
\begin{equation}\label{selectionrulescase(a)}
\Delta L = \pm 1,\,\,\,\,\,\,\,\, \Delta M_{L} = 0,
\end{equation}
together with a change in total parity $\pi$.\ Note that we have assumed an initial atomic state with $M_{L}=0$ (for example, the $S_{0}^{e}$ ground-state), such that $L$-conserving transitions are forbidden.\  As a result of the aforementioned selection rules, the Hamiltonian matrix $(H_{kk'})$ exhibits the following block-tridiagonal structure,
\begin{equation} \label{linpolham}
(H_{kk'}) =
\begin{pmatrix}
 H_{ {S_{0}^{e}}{S_{0}^{e}} }& H_{ {S_{0}^{e}}{P_{0}^{o}} } & 0 & 0 & 0 & \cdots &  \\ 
 H_{ {P_{0}^{o}}{S_{0}^{e}} }& H_{ {P_{0}^{o}}{P_{0}^{o}} } & H_{ {P_{0}^{o}}{D_{0}^{e}} } & 0 & 0 & \cdots \\
 0 & H_{ {D_{0}^{e}}{P_{0}^{o}} }   & H_{ {D_{0}^{e}}{D_{0}^{e}} } & H_{ {D_{0}^{e}}{F_{0}^{o}} } & 0 & \cdots \\
 0 & 0 & H_{ {F_{0}^{o}}{D_{0}^{e}} } & H_{ {F_{0}^{o}}{F_{0}^{o}} } & H_{ {F_{0}^{o}}{G_{0}^{e}} } & \cdots \\
 0 & 0 & 0 & H_{ {G_{0}^{e}}{F_{0}^{o}} } & H_{ {G_{0}^{e}}{G_{0}^{e}} } & \cdots \\
\vdots & \vdots & \vdots & \vdots & \vdots & \ddots
\end{pmatrix}.
\end{equation}
Here, we have adopted the notation $L_{M_{L}}^{\pi}$ for the various $(N+1)$-electron target states, omitting the spin multiplicity (which, in the present non-relativistic theory, is conserved in any transition).\ The diagonal blocks are, individually, diagonal matrices, consisting of the eigenvalues of the Hamiltonian operator with respect to the basis of field-free eigenfunctions $\psi_{k}({\bf X}_{N+1})$.\ Their dimensions are determined by the total number of $(N + 1)$-electron configurations that admit those angular symmetry properties.\ The off-diagonal blocks consist of the dipole matrix elements.\ It should be emphasized that the linearly polarized laser field couples each target state to no more than two others (for initial $M_{L}=0$), so that the bandwidth of the matrix (\ref{linpolham}) never spans more than a single block.\ This represents a particularly simple structure for numerical computations.

Of course, in any practical calculation, only a finite number of basis functions $\psi_{k}({\bf X}_{N+1})$ can be included.\ The basis set is rendered finite by imposing an upper limit, $L_{\textrm{max}}$, on the total angular momentum $L$ of the $(N + 1)$-electron states.\ As a result, the number of target $LS\pi$ symmetries, in the present case (with selection rules given by Eqs.\  (\ref{selectionrulescase(a)})), is restricted to
\begin{equation}\label{Nsymlin}
N_{\textrm{sym}} = L_{\textrm{max}} + 1.
\end{equation}
The choice of $L_{\textrm{max}}$, in turn, is largely dictated by the radiation-field parameters in the problem of interest.\ In particular, the number of target angular momenta (and therefore basis-set size), required for numerical convergence, scales rapidly with wavelength \cite{Hassounehetal2015,Hassounehetal2014}.\ As a result, the study of strong-field processes in the long-wavelength optical and near-infrared regimes can become prohibitively demanding, necessitating an efficient computational strategy for both storing, and performing calculations with, large Hamiltonian matrices.

Additionally, we highlight that a substantial increase in the dimensions of $(H_{kk'})$ can also occur in treating more general initial states of the target \cite{Hutchinsonetal2011}.\ For atomic systems with aligned initial states (e.g., Ne$^{+}$ or Ar$^{+}$, in their $P^{o}$ ground state, with $M_{L}= 1$), the selection rule on $L$ is relaxed to $\Delta L = 0, \pm 1$, whilst $M_{L}$ remains conserved, $\Delta M_{L} = 0$.\ The possibility of transitions with $\Delta L = 0$  increases the number of accessible symmetries, allowing both even and odd parity for each orbital angular momentum of the $(N + 1)$-electron system.\ Moreover, the number of dipole-couplings is enhanced, with each $LS\pi$ state of the target interacting with up to three others (rather than two in the case of $M_{L} = 0$).\ Such conditions further stress the need for an efficient scheme of computation in time-dependent $R$-matrix approaches, accommodating large quantities of atomic structure and dipole-coupling data.

The current implementation of time-dependent $R$-matrix theory, in the form of the RMT approach, provides an efficient means of treating problems in which the $R$-matrix basis must be enlarged, whether due to a change in the selection rules (for aligned target states), or a more demanding set of conditions for numerical convergence.\ Here, application of the Arnoldi algorithm \cite{Mooreetal2011} enables the numerical solution of Eq.\ (\ref{ODEs}) through a series of matrix-vector multiplications.\ Not only does this facilitate accurate propagation of the wavefunction, but the memory demands, imposed by the inclusion of more target symmetries, can be mitigated through a block-distribution of the matrix, and vector, across multiple parallel processors.\ Presently, the RMT suite of codes exploit the message passing interface (MPI) library to achieve this data decomposition.\ Each target $LS\pi$ symmetry is assigned to one or more MPI tasks, and to facilitate the calculation of matrix-vector products arising in the Arnoldi method, blocks of the wavefunction vector are sent and received dynamically (during each time-step of the propagation).\ The computational tractability afforded by the RMT method, when a large number of target symmetries and their dipole-coupling need to be accounted for, renders it suitable for subsequent developments and still more demanding applications, particularly in regard of arbitrarily polarized light fields.\ Indeed, prior to the progress reported here, both the data distribution, and parallel communication strategies, implemented in the RMT codes were uniquely specialized to the Hamiltonian structure (\ref{linpolham}) and its analogue for aligned initial states.

{\it (b) Arbitrarily polarized laser field:} The interaction of a neutral, noble gas atom with an arbitrarily polarized radiation field naturally presents the greatest complexities.\ When all three components of the electric field (\ref{field}) are active, the dipole selection rules become
\begin{equation}\label{selectionrulesarbfield}
\Delta L = 0, \pm 1,\,\,\,\,\,\,\,\, \Delta M_{L} = 0, \pm 1,
\end{equation}
in addition to a change in parity.\ Correspondingly, the Hamiltonian matrix assumes the form
\begin{widetext}
\begin{equation}\label{Hamarb}
(H_{kk'}) =
\begin{pmatrix}

H_{ {S_{0}^{e}}{S^{e}_{0}} } & 0 & 0 & 0 & H_{ {S_{0}^{e}}{P_{-1}^{o}} } & H_{ {S_{0}^{e}}{P_{0}^{o}} } & H_{ {S_{0}^{e}}{P_{1}^{o}} } & \cdots \\

 0 & H_{ {P_{-1}^{e}}{P_{-1}^{e}} } & 0 & 0 & H_{ {P_{-1}^{e}}{P_{-1}^{o}} } & H_{ {P_{-1}^{e}}{P_{0}^{o}} } & 0 & \cdots \\

 0 & 0 & H_{ {P_{0}^{e}}{P_{0}^{e}} } & 0 & H_{ {P_{0}^{e}}{P_{-1}^{o}} } & H_{ {P_{0}^{e}}{P_{0}^{o}} } & H_{ {P_{0}^{e}}{P_{1}^{o}} } & \cdots \\

 0 & 0 & 0 & H_{ {P_{1}^{e}}{P_{1}^{e}} } & 0 & H_{ {P_{1}^{e}}{P_{0}^{o}} } & H_{ {P_{1}^{e}}{P_{1}^{o}} } & \cdots \\

H_{ {P_{-1}^{o}}{S_{0}^{e}} } & H_{ {P_{-1}^{o}}{P_{-1}^{e}} } & H_{ {P_{-1}^{o}}{P_{0}^{e}} } & 0 & H_{ {P_{-1}^{o}}{P_{-1}^{o}} } & 0 & 0 &\cdots \\

H_{ {P_{0}^{o}}{S_{0}^{e}} } & H_{ {P_{0}^{o}}{P_{-1}^{e}} } & H_{ {P_{0}^{o}}{P_{0}^{e}} } & H_{ {P_{0}^{o}}{P_{1}^{e}} } & 0 & H_{ {P_{0}^{o}}{P_{0}^{o}} } & 0 &  \cdots \\

H_{ {P_{1}^{o}}{S_{0}^{e}} } & 0 & H_{ {P_{1}^{o}}{P_{0}^{e}} } & H_{ {P_{1}^{o}}{P_{1}^{e}} } & 0 & 0 & H_{ {P_{1}^{o}}{P_{1}^{o}} } & \cdots \\

\vdots & \vdots & \vdots & \vdots & \vdots & \vdots & \vdots & \ddots

\end{pmatrix}.
\end{equation}
\end{widetext}
This Hamiltonian governs a richer dynamics than that of Eq.\ (\ref{linpolham}), by virtue of the increased number of relevant electronic degrees of freedom.\ Computationally, the difficulties arising from its treatment are twofold.\ First, the replacement of each $LS\pi$ symmetry, with $(2L + 1)$ $LM_{L}S\pi$ symmetries, incurs a dramatic increase in the size of the matrix.\ Specifically, the number of symmetries (diagonal blocks), for a given choice of $L_{\textrm{max}}$, is now given by
\begin{equation}\label{Nsym}
N_{\textrm{sym}} = 2(L_{\textrm{max}} + 1)^{2} - 1.
\end{equation}
Subtraction of unity in this equation accounts for the absence of the $S_{0}^{o}$ symmetry, which is dipole-inaccessible for neutral, noble-gas systems.\
Thus, when all possible magnetic substates are accounted for explicitly, $N_{\textrm{sym}}$ scales quadratically with $L_{\textrm{max}}$.\ This contrasts with the linear scaling of Eq.\ (\ref{Nsymlin}) in the case of pure, linear polarization.\ Second, whenever $\Delta M_{L} = 0, \pm 1$ transitions are permitted, the total number of dipole-couplings is enhanced.\ This is reflected by an increase in the number of off-diagonal (dipole) blocks, which are no longer distributed in the simple manner of (\ref{linpolham}) (along the block super- and sub-diagonals), but which now span a much larger bandwidth of the matrix.\ As a result, it becomes essential to manage a much more intricate set of parallel communications, among MPI tasks responsible for different target symmetries, whenever the matrix-vector multiplications are performed in a block-distributed fashion.

To tackle these complications, and thereby extend the predictive capabilities of the RMT method to include arbitrary light fields, we have made several critical modifications to the suite of codes.\ Now, each $LM_{L}S\pi$ (as opposed to $LS\pi$) symmetry is assigned to one or more MPI tasks, so that Eq.\ (\ref{Nsym}) (rather than Eq.\ (\ref{Nsymlin})) constitutes the minimum number of processor cores required for the inner-region computations.\ Such a scheme suffices for the description of atomic ionization in low-intensity, XUV laser fields, for which only a limited number of angular momenta $(L_{\textrm{max}} \approx 10)$ are required for satisfactory convergence (see Sections {\ref{sec:He}} and {\ref{sec:Fminus}}).\ However, as suggested by our previous work \cite{Hassounehetal2015,Hassounehetal2014}, the study of strong-field processes in long-wavelength (especially optical and near-infrared) fields necessitates much larger values of $L_{\textrm{max}}$ for good convergence $(L_{\textrm{max}} \approx 100\,{\textendash}\,200)$.\ To render such problems tractable, we have implemented a number of computational measures and simplifications, which reduce core requirements and improve load balancing for the inner-region calculations.\ We mention two in particular.\ Firstly, a parameter $M_{L}^{\textrm{max}}$ has been introduced, which limits the target magnetic substates to a range $\vert M_{L}\vert \leq M_{L}^{\textrm{max}}$.\ This parameter proves valuable when only a subset of these are significantly populated during the dynamics.\ Such behaviour is realized, for instance, in the cross-polarized laser-field configurations explored in recent two-color HHG experiments \cite{Lambertetal2015,Soiferetal2013}, where $\Delta M_{L} = \pm 1$ transitions are minimized through an appropriate choice of $z$-axis (i.e., such that it coincides with the polarization axis of the longer wavelength, and/or higher intensity, laser pulse).\ Under such conditions, restricting the number of magnetic substates so that $M_{L}^{\textrm{max}} \ll L_{\textrm{max}}$ can facilitate a substantial reduction in the number of $LM_{L}S\pi$ symmetries retained in the calculations, now given by
\begin{align}
N_{\textrm{sym}} = &\, 2[(M_{L}^{\textrm{max}}+1)^{2} \nonumber \\
& + (2M_{L}^{\textrm{max}}+1)(L_{\textrm{max}} - M_{L}^{\textrm{max}})]-1 \nonumber
\end{align}
instead of Eq.\ (\ref{Nsym}), and thus, in the number of processor cores assigned to the inner-region.\ Secondly, whilst we must set $M_{L}^{\textrm{max}} = L_{\textrm{max}}$ for linearly or elliptically polarized fields in the $xy$-plane, not all $LM_{L}S\pi$ symmetries of the target are realizable via dipole transitions.\ Indeed, when only the $x$- and/or $y$-components of the electric field (\ref{field}) assume non-zero values, the selection rule on $M_{L}$ is $\Delta M_{L}=\pm 1$.\ Then, for a neutral, noble-gas system, irradiated by a linearly or circularly polarized field in the $xy$-plane (see, for example, Section {\ref{sec:He}}), the symmetries $P^{e}_{\pm 1}$, $P^{o}_{0}$, $D^{e}_{\pm 1}$, $D^{o}_{0}$, $D^{o}_{\pm 2}, ...$ are all inaccessible by dipole-allowed transitions from the $S^{e}_{0}$ ground state.\ In practice, we therefore exclude the corresponding symmetry blocks from the Hamiltonian (\ref{Hamarb}) when such fields are considered.\ This affords a considerable saving in computational effort, for the number of target $LM_{L}S\pi$ symmetries is almost halved relative to (\ref{Nsym}),
\begin{equation}
N_{\textrm{sym}} = (L_{\textrm{max}} + 1)^{2}. \nonumber
\end{equation}
Such measures to limit the computational load, where possible, aid in expanding the range of intensities and wavelengths which can be explored using the latest RMT code for arbitrarily polarized fields.

In tandem with these modifications, we have also adapted the parallel linear algebra routines, critical to the implementation of the Arnoldi propagator, for the Hamiltonian (\ref{Hamarb}).\ To compute matrix-vector products involving this matrix, blocks of the vector must be sent and received among MPI tasks responsible for different target symmetries.\ Parallel communication strategies, employed in the original RMT codes for linearly polarized fields, were developed specifically for the block-tridiagonal structure of (\ref{linpolham}), as well as its analogue in the case of aligned initial states $(M_{L} \neq 0)$.\ To accomodate the Hamiltonian (\ref{Hamarb}), relevant for a field of arbitrary polarization, we have devised a much more robust set of communication routines for the efficient exchange of data.\ Crucially, our scheme is now based solely on the $LM_{L}S\pi$ couplings prevalent in the problem of interest, and not on a fixed structure of the Hamiltonian matrix (that is, a specific arrangement of the off-diagonal dipole blocks).\ This has two important implications.\ Firstly, a high degree of efficiency is maintained, by avoiding the unnecessary sending or receiving of data whenever only a subset of the selection rules (\ref{selectionrulesarbfield}) are satisfied.\ This is valuable in particular special cases of (\ref{Hamarb}) (e.g., for an elliptically polarized field in the $xy$-plane, $\Delta M_{L} = 0$ transitions are forbidden, and the corresponding dipole matrix elements are zero).\ Secondly, our scheme could be adapted to manage the communications required for other interactions, such as those of a non-dipole or relativistic nature.

\subsection{Outer Region}
\label{subsec:Outer}

To solve (\ref{TDSE}) in the outer region, we expand the time-dependent, $(N + 1)$-electron wavefunction in a standard close-coupling form,
\begin{equation}\label{RMTOuter}
\Psi\left({\bf X}_{N+1},t\right) = \sum_{p}\bar{\Phi}_{p}\left({\bf X}_{N};{\hat{\bf r}}_{N+1}\sigma_{N+1}\right)\frac{1}{r}f_{p}(r,t).
\end{equation}
Here, the radial coordinate of the ejected electron, $r_{N+1}$, is denoted as $r$ for brevity.\ The channel functions $\bar{\Phi}_{p}\left({\bf X}_{N};{\hat{\bf r}}_{N+1}\sigma_{N+1}\right)$ are formed by coupling the orbital and spin angular momenta of the residual ion with those of the outgoing electron \cite{BurkeBerrington1993,Burke2011}.\ The time-dependence of the multielectron wavefunction is incorporated solely in the functions $f_{p}(r,t)$, which describe the radial motion of the ejected electron in each channel $p$.\ Note that expansion (\ref{RMTOuter}) is unsymmetrized:\ the spatial isolation of the ionized electron, from the complex, many-body inner region, ensures that the exchange interaction is negligible.\ Moreover, since the number of electrons in the outer region is limited to one, the dimensionality of the TDSE, for each residual-ion state, is reduced to at most three.\ This affords a substantial simplification of the computational problem.\

As in the original formulation of RMT theory \cite{Mooreetal2011}, it can be shown that the one-electron, reduced radial wavefunctions $f_{p}(r,t)$ satisfy a system of coupled, second-order, partial differential equations,
\begin{align}\label{PDEs}
i\frac{\partial}{\partial t}f_{p}(r,t) =  h_{p}(r)f_{p}(r,t) & + \sum_{p'}\bigl[W_{pp'}^{E}(r) + W_{pp'}^{D}(t) \nonumber \\
& +  W_{pp'}^{P}(r,t)\bigr]f_{p'}(r,t).
\end{align}
The one-electron operator $h_{p}(r)$, given by
\begin{equation}
h_{p}(r) = -\frac{1}{2}\frac{d^{2}}{dr^{2}} + \frac{l_{p}(l_{p}+1)}{2r^{2}} - \frac{Z-N}{r} + E_{p}, \nonumber
\end{equation}
includes terms corresponding to the kinetic energy, screened nuclear attraction and centrifugal repulsion for the ejected electron.\ The quantities $l_{p}$ and $E_{p}$ are the angular momentum of the outgoing electron, and the energy of the residual-ion state, respectively.\ The remaining terms on the right-hand side of Eq.\ (\ref{PDEs}) correspond to the long-range potentials \cite{Mooreetal2011,Lysaghtetal2009}.\ The matrix $W^{E}$ has been discussed in the context of time-independent formulations of $R$-matrix theory \cite{BurkeBerrington1993,Burke2011}, and arises from the repulsive interaction among the outgoing and residual electrons,
\begin{equation}
W_{pp'}^{E}(r) = \left\langle \bar{\Phi}_{p} \left\vert \sum_{j=1}^{N}\frac{1}{\vert {\bf r} - {\bf r}_{j}\vert} - \frac{N}{r} \right\vert \bar{\Phi}_{p'}\right\rangle . \nonumber
\end{equation}
The matrix $W^{D}$ is time-dependent, and describes the interaction of the light field with the $N$-electron residual ion,
\begin{equation}\label{DEFWD}
W_{pp'}^{D}(t) = \left\langle \bar{\Phi}_{p}\lvert {\bf E}(t)\cdot {\bf R}_{N}\rvert \bar{\Phi}_{p'} \right\rangle,
\end{equation}
where ${\bf R}_{N}$ is given by the sum of the position operators ${\bf r}_{i}$ for electrons $i = 1,..,N$.\ Finally, $W^{P}$ emerges from the interaction of the light field with the ejected electron,
\begin{equation}\label{DEFWP}
W_{pp'}^{P}(r,t) = \left\langle \bar{\Phi}_{p} \left\vert {\bf E}(t)\cdot {\bf r}\right\vert \bar{\Phi}_{p'}\right\rangle.
\end{equation}
Note that the integration, implied in each of these equations, is performed over all electron space and spin coordinates, with the exception of the radial coordinate of the ejected electron.

The field-dependent potentials (\ref{DEFWD}) and (\ref{DEFWP}) play a critical role in the RMT outer-region analysis.\ Information pertaining to the polarization of the laser field, and concomitantly, the dipole selection rules for laser-induced transitions, is encoded entirely therein.\ Previously, these had been derived, and implemented, purely for linearly polarized fields (in the $z$-direction), so that the couplings, among different electron emission channels, were appropriate only for $\Delta M_{L} = 0$ transitions of the $(N+1)$-electron system.\ To enable the treatment of light fields with arbitrary polarization, we have established a more general set of potentials, which also express the essential channel couplings when $\Delta M_{L} = \pm 1$ transitions are allowed.\ Their derivation generalizes that given in Ref.\ \cite{Lysaghtetal2009} for fields linearly polarized in the $z$-direction, and relies on standard techniques for irreducible tensor operators \cite{Varshalovichbook}.\ We summarize the results here for reference,
\begin{equation}\label{WDsum}
W_{pp'}^{D} = \sum_{\mu = -1,0,1}W_{pp'}^{D\,(\mu)},
\end{equation}
\begin{align}
W_{pp'}^{D\,(\mu)} = \,\, & (-1)^{L_{p}+L_{p'}+\mu}E_{\mu}(t) \sqrt{2L'+1} \nonumber \\
& \times (1(-\mu)L'M_{L'}\vert LM_{L})W(1L_{p'}Ll_{p};L_{p}L') \nonumber \\ & \times \langle \alpha_{p}L_{p} \vert\vert R_{N}\vert \vert \alpha_{p'}L_{p'}\rangle \nonumber \\
& \times \delta_{l_{p}l_{p'}}\delta_{m_{l_{p}}m_{l_{p'}}}\delta_{SS'}\delta_{S_{p}S_{p'}}\delta_{M_{S}M_{S'}}\delta_{M_{S_{p}}M_{S_{p'}}},
\end{align}
and
\begin{equation}
W_{pp'}^{P} = \sum_{\mu = -1,0,1} W_{pp'}^{P\,(\mu)},
\end{equation}
\begin{align}\label{WPexpression}
W_{pp'}^{P\,(\mu)} = &\,\, (-1)^{L+L'+1}E_{\mu}(t) \sqrt{(2l_{p'}+1)(2L+1)} \nonumber \\
& \times (LM_{L}1\mu\vert L'M_{L'})W(1l_{p'}LL_{p};l_{p}L') \nonumber \\
& \times \frac{a(l_{p'})}{(l_{p}010\vert l_{p'}0)}r \nonumber \\
& \times \delta_{\alpha_{p}\alpha_{p'}}\delta_{L_{p}L_{p'}}\delta_{SS'}\delta_{S_{p}S_{p'}}\delta_{M_{S}M_{S'}}\delta_{M_{S_{p}}M_{S_{p'}}}.
\end{align}
In these equations, $E_{\mu}(t)$ are the spherical components of the electric field intensity.\ For a given channel $p$, $L, M_{L}$ and $S, M_{S}$ denote the total orbital and spin angular momentum quantum numbers, whilst $L_{p}, M_{L_{p}}$ and $S_{p}, M_{S_{p}}$ are the quantum numbers pertaining to the residual ion state.\ Also, $l_{p}$ and $m_{l_{p}}$ are the orbital angular momentum quantum numbers associated with the ejected electron.\ All remaining quantum numbers, required to specify the ionic state, are denoted collectively by $\alpha_{p}$.\ The quantities $\langle \alpha_{p}L_{p} \vert\vert R_{N}\vert \vert \alpha_{p'}L_{p'}\rangle$ are the reduced matrix elements of the $N$-electron position operator.\ Finally, $a(l_{p'})$ is defined by
\begin{equation}
a(l_{p'})=\begin{cases}
\frac{\displaystyle{l_{p'}}}{\displaystyle{[(2l_{p'} - 1)(2l_{p'} + 1)]^{1/2}}},\,\,\, l_{p} = l_{p'} - 1 \\
\\
-\frac{\displaystyle{(l_{p'}+1)}}{\displaystyle{[(2l_{p'} + 1)(2l_{p'} + 3)]^{1/2}}},\,\,\, l_{p} = l_{p'} + 1. \\
\end{cases} \nonumber
\end{equation}
Throughout, we have employed the Fano-Racah phase convention \cite{FanoRacah}.\ It should be noted that, in the case of a field linearly polarized along the $z$-axis $(E_{\pm 1} = 0)$, we recover the potentials employed in previous formulations of time-dependent $R$-matrix theory \cite{Mooreetal2011,Lysaghtetal2009,Burke2011}.\

\section{Application to Two-photon Ionization of H\lowercase{e} in Circularly Polarized Light Fields}
\label{sec:He}

As a first demonstration of the generalized RMT methodology, we investigate the formation of multistart, spiral vortex features in the photoelectron momentum distributions of He, irradiated by a pair of time-delayed, ultrashort, circularly polarized laser pulses with opposite helicities.\ We validate our results through comparison of the RMT data with that of Ngoko Djiokap {\it et al.} \cite{Djiokapetal2016}, who previously treated the same laser-driven, two-electron problem by means of the TDCC approach \cite{Pindzolaetal2007,ColganPindzola2012} in conjunction with Wigner frame transformations \cite{Muller1997,Kjeldsenetal2007}.\

\subsection{Calculation Parameters}
\label{subsec:paramHe}

The He target considered in this work is as discussed in previous $R$-matrix studies \cite{Hutchinsonetal2010,BrownvanderHart2012,ReyvdH2014}.\ Within the inner region, we regard the atomic system as He$^{+}$ to which is added a single electron.\ For the description of He$^{+}$, we employ the physical $1s$ orbital, together with $\overline{2s}$ and $\overline{2p}$ pseudo-orbitals.\ The pseudo-orbitals are expressed analytically as a sum of Sturmian-type orbitals, each of the form $r^{i}e^{-\alpha r}$, with the same exponential decay as the $1s$ function, minimal degree of the polynomial, and orthogonality with respect to the $1s$ function.\ Their inclusion facilitates a more accurate account of changes in the He$^{+}$ ground state, induced by the laser field, than might be achieved with the physical orbitals alone.\ The initial state is the He $1s^{2} \, {^{1}S^{e}}$ ground state, with binding energy $E_{b}(\textrm{He}) \approx 24.6\textrm{ eV}$.\

The radial extent of the inner-region is $20\textrm{ a.u.}$, which suffices to effectively confine the orbitals of the residual He$^{+}$ ion.\ The inner-region continuum functions are generated using a set of 70 $B$-splines of order 9 for each available orbital angular momentum of the outgoing electron.\ The knot-point distribution varies from a near-quadratic spacing, in proximity to the nucleus, to a near-linear spacing towards the inner-region boundary.\ Additional knot points are inserted, further inward, to improve the description of the one-electron continuum functions close to the nucleus.\ We retain all admissible electron emission channels up to a maximum total orbital angular momentum $L_{\textrm{max}} = 9$, as well as all permitted magnetic substates with $-9 \leq M_{L} \leq 9$.\ The outer-region boundary radius is $3500\textrm{ a.u.}$, ensuring that no unphysical interference structure in the wavefunctions arise through reflection of the ejected electron wavepacket.\ The finite-difference grid spacing is $0.08\textrm{ a.u.}$.\ To advance the multielectron wavefunction in time, we adopt an Arnoldi propagator of order 8, choosing a time-step of $0.01\textrm{ a.u.}$.

We select a set of pulse characteristics appropriate for the single-color, two-photon interferometry scheme proposed by Ngoko Djiokap {\it et al.}\ \cite{Djiokapetal2016}.\ Therein, the He atom is subject to a pair of counter-rotating, circularly polarized, femtosecond laser pulses, having controlled time-delay $\tau$, and CEP values $\varphi_{1}$ and $\varphi_{2}$.\ Both pulses exhibit a sine-squared ramp-on/off temporal profile.\ For such a configuration, the electric field may be expressed in the form
\begin{align}\label{TwopulsesHe}
{\bf E}(t) = & F(t)\textrm{Re}[ {\bf e}_{1}\,e^{-i(\omega t + \varphi_{1})}] \nonumber \\
& + F(t - \tau)\textrm{Re}[ {\bf e}_{2}\,e^{-i(\omega (t - \tau) + \varphi_{2})}],
\end{align}
where, for circular polarization in the $xy$-plane, ${\bf e}_{1} = {\bf e}_{2}^{*} = ( \hat{\bf x} + i\hat{\bf y} )/\sqrt{2}$.\ The sine-squared envelope is described by
\begin{equation}\label{sinesquared}
F(t) = F_{0}\sin^{2}(\omega t/2N),
\end{equation}
with $F_{0}$ the peak electric field intensity.\ The latter is related to the pulse peak intensity $I_{0}$ by $I_{0} = (c/4\pi)F_{0}^{2}$, where $c$ is the speed of light in vacuum.\ In line with Ngoko Djiokap {\it et al.}\ \cite{Djiokapetal2016}, we adopt a carrier frequency of $\omega = 15\textrm{ eV}$, a peak intensity of $I_{0} = 1\times 10^{12}\,\textrm{Wcm}^{-2}$, and a duration of $N = 6$ cycles for each pulse.\ Note that the time-delay is measured between the central peaks of the two pulses, and is always chosen such that the right-hand circularly polarized pulse attains peak intensity simultaneously with, or before, its left-hand circularly polarized counterpart (i.e., $\tau \geq 0$ in Eq.\ (\ref{TwopulsesHe})).

By solving the TDSE, it is well-known \cite{Madsenetal2007} that the ionized-electron momentum distribution can be extracted from the wavepacket solution in three different zones:\ (i) reaction zone, (ii) Coulomb zone, and (iii) free zone.\ In the reaction zone, the momentum distribution is obtained by projecting the wavepacket (immediately after termination of the pulse) onto fully correlated, field-free, scattering wavefunctions.\ In Ref.\ \cite{Djiokapetal2016}, the latter were computed by the $J$-matrix method \cite{Djiokapetal2012}.\ In the Coulomb zone, the photoelectron properties are assessed through projection of the ionized-electron wavepacket (long after termination of the laser pulse) onto field-free, continuum wavefunctions, approximated by a product of a Coulomb function and a bound-state wavefunction.\ In the TDCC study of Ngoko Djiokap {\it et al.}\ \cite{Djiokapetal2016}, such projections were performed at times up to $20\textrm{ a.u.}$ following the end of the laser pulse, with good agreement observed between these Coulomb-zone spectra and that obtained in the reaction zone.\ Finally, in the free zone, the momentum distribution is calculated by projecting the wavepacket (after a substantially long period of time following the end of the pulse) onto a product of a plane-wave wavefunction and a bound-state wavefunction.\ This method, which requires extremely large simulation domains, is equivalent to a Fourier transform of the wavepacket at long times, and is akin to the procedure adopted in our present investigation.

To ascertain photoelectron momentum spectra in the RMT approach, we must determine the angular momentum characteristics of the continuum electron wavefunction in each ionization channel of interest.\ Following a time-dependent simulation, we obtain data for the reduced radial wavefunctions (denoted $f_{p}$ in Eq.\ (\ref{RMTOuter})), for every channel, at the final time-step.\ However, all information pertaining to the orbital and spin angular momenta of the outgoing electron is associated with the channel functions (denoted $\bar{\Phi}_{p}$ in Eq.\ (\ref{RMTOuter})).\ To extract the spatial component of the continuum electron wavefunction (including the angular dependence) in each channel, we decouple the orbital and spin angular momenta of the ejected electron and residual He$^{+}$ states, employing Clebsch-Gordan coefficients \cite{vanderHartLysaghtBurke2008}.\ Once acquired, we transform the wavefunction, for $r > 200\textrm{ a.u.}$, into the momentum representation, under the assumption that the long-range Coulomb potential is negligible.

It is pertinent to highlight three important facets of this analysis in the present context.\ Firstly, although we retain the non-physical $\overline{2s}$ and $\overline{2p}$ pseudo-thresholds in computing the momentum spectra, their associated ionization channels are populated insignificantly under the prevalent field conditions.\ This is to be expected:\ whilst absorption of at least two $15\textrm{-eV}$ photons is required for production of the He$^{+}$ $1s$ state, the $\overline{2s}$ and $\overline{2p}$ pseudo-thresholds are accessible only through absorption of at least five such photons.\ The low pulse intensities, regarded in this work, ensure that such higher-order processes occur with negligible probability.\ As a result, the presence of the pseudo-thresholds has no adverse consequences for our comparison with the data of Ngoko Djiokap {\it et al.}\ \cite{Djiokapetal2016}, who project the full, two-electron wavefunction onto field-free, scattering wavefunctions of the singly-ionized $\textrm{He}^{+}(1s) + e^{-}$ continuum, constructed by means of the $J$-matrix method \cite{Djiokapetal2012}.\ Secondly, to minimize the role of dielectronic repulsion, and thus ensure the validity of our procedure for assessing photoelectron emission properties, we have propagated the total wavefunction for a further 116 field cycles following termination of the second laser pulse.\ Additional cycles of field-free propagation (up to a total of 300) incur no significant alterations to the spectra.\ Thirdly, a minimum cutoff distance of $200\textrm{ a.u.}$ was deemed suitable following a direct examination of the ejected-electron wavefunctions.\ Beyond this distance, the latter exhibit clear continuum-wave characteristics.\ We have also repeated our analysis for other distances, verifying that the main features of the spectra are both qualitatively, and quantitatively, insensitive to acceptable variations of the minimum cutoff.\ Note, however, that effecting the transformation in this manner does artificially eliminate near-zero-momentum features of the distribution, which arise from threshold photoelectrons.\ The latter are typically 10 times weaker than the dominant emission features discussed in Section {\ref{subsec:HeResults}}, and are not of interest for the present study.

\subsection{Results}
\label{subsec:HeResults}

Figures {\ref{fig:Fig1}} and {\ref{fig:Fig2}} present photoelectron momentum distributions, in the polarization plane, for selected CEPs and time-delays between the pulses.\ Our numerical RMT results (Figures {\ref{fig:Fig1}}(a),(c) and {\ref{fig:Fig2}}(a),(c),(e)) are compared with the TDCC data reported by Ngoko Djiokap {\it et al.}\ \cite{Djiokapetal2016} (Figures {\ref{fig:Fig1}}(b),(d) and {\ref{fig:Fig2}}(b),(d),(f)).\ Note that, in Ref.\ \cite{Djiokapetal2016}, only the two-photon ionization channels (i.e., those with $L = 0,2$ and $M_{L} = 0, \pm 2$) were retained in computing the momentum maps of Figures {\ref{fig:Fig1}}(b),(d) and {\ref{fig:Fig2}}(b),(d),(f).\ To enable a true comparison, we also include exclusively these channels in our numerical projections.\ In any case, our data suggests that these channels account for around $90\%$ of the total ionization yield.

\begin{figure}
\subfloat{
  \includegraphics[clip,height = 0.235\textwidth,width=0.235\textwidth]{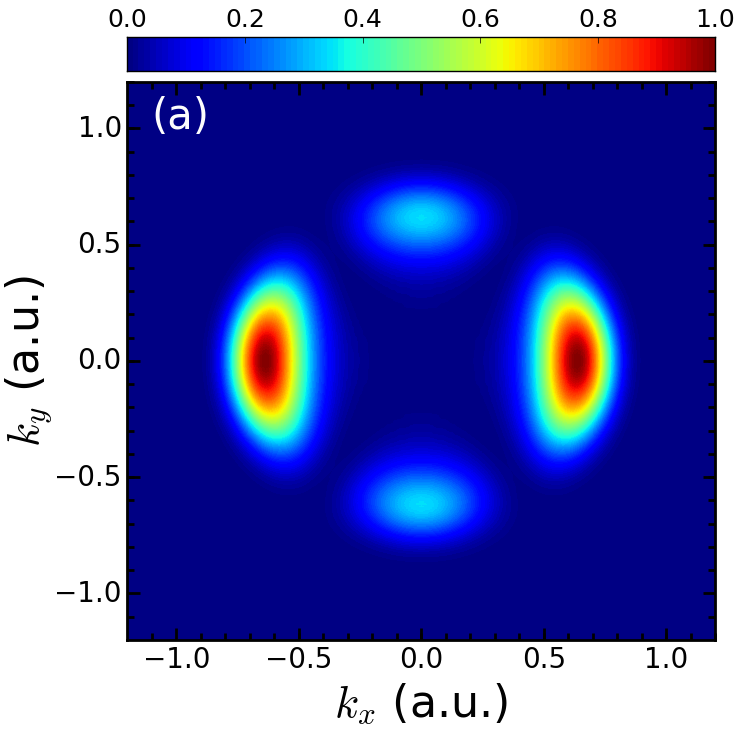}
}
\subfloat{
  \includegraphics[clip,height = 0.235\textwidth,width=0.235\textwidth]{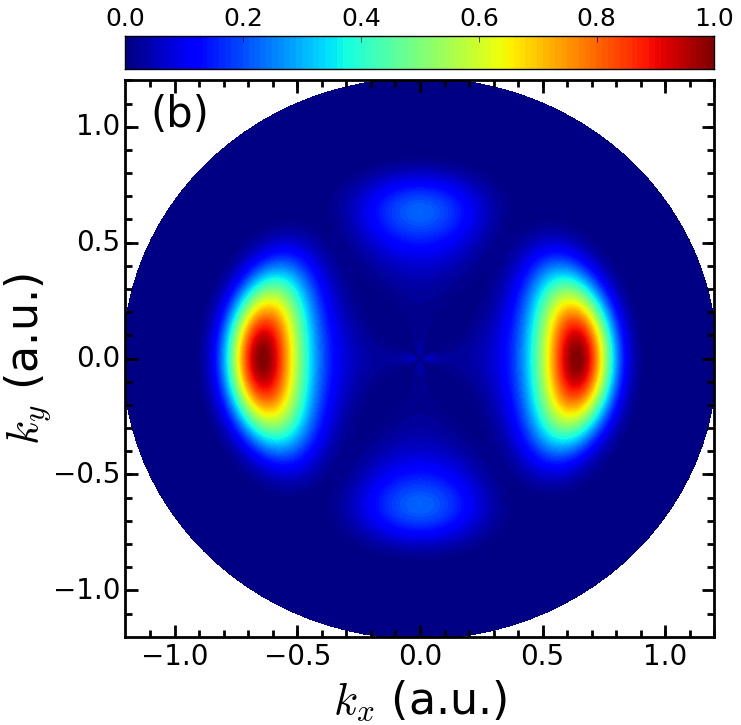}
}

\subfloat{
  \includegraphics[clip,height = 0.235\textwidth,width=0.235\textwidth]{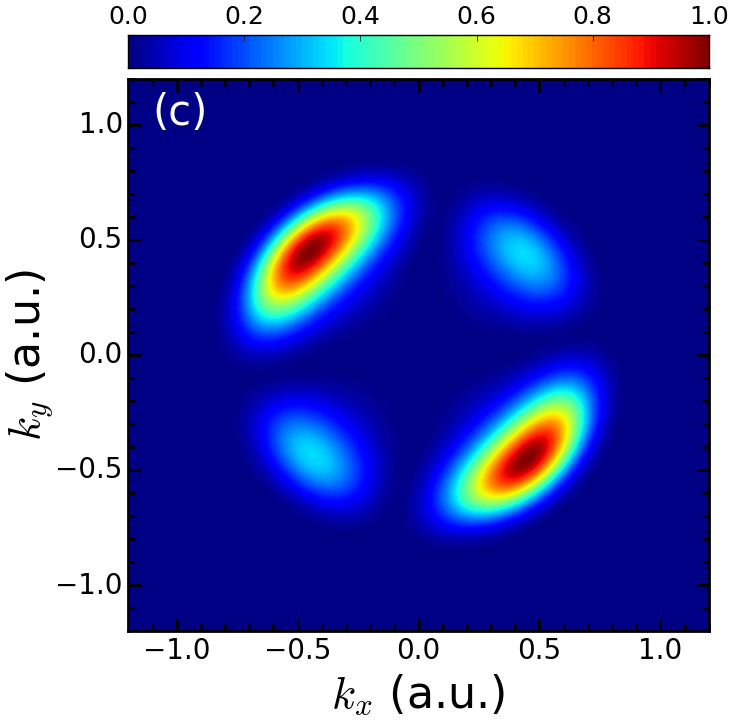}
}
\subfloat{
  \includegraphics[clip,height = 0.235\textwidth,width=0.235\textwidth]{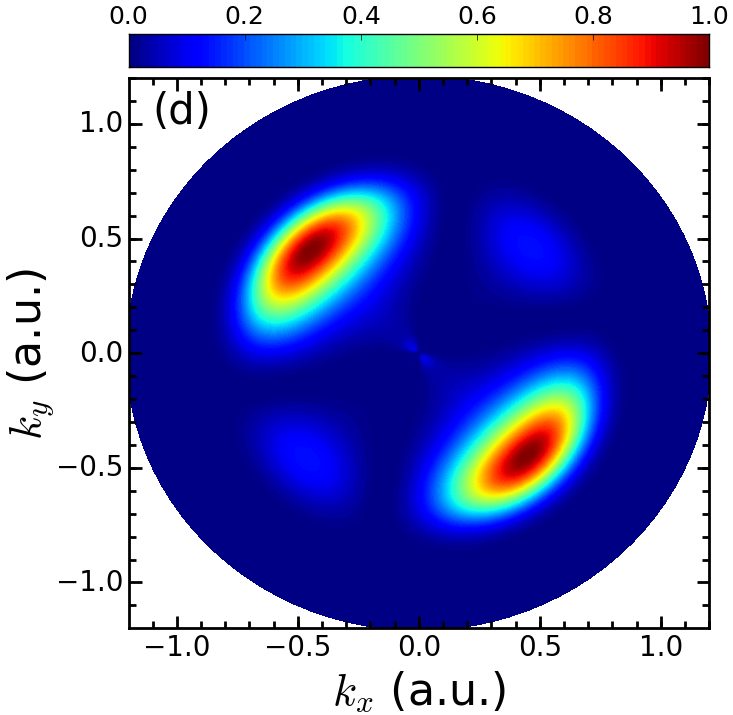}
}
\caption{(Color online) Photoelectron momentum distributions, in the polarization plane, following ionization of He by right- and left-hand circularly polarized laser pulses with zero time-delay $(\tau = 0)$.\ The distributions acquired through the generalized RMT method (for arbitrary polarization) are shown in (a) and (c), whilst those obtained by Ngoko Djiokap {\it et al.}\ \cite{Djiokapetal2016}, employing a TDCC approach, are shown in (b) and (d).\ Each pulse has a carrier frequency of $15\textrm{ eV}$, a duration of 6 cycles and a peak intensity of $1\times 10^{12}\,\textrm{Wcm}^{-2}$.\ In (a) and (b), the CEPs are $\varphi_{1} = \varphi_{2} = 0$, but in (c) and (d), they are chosen to be $\varphi_{1} = 0, \varphi_{2} = \pi/2$.\ Relative magnitudes are indicated by the color scales.}
\label{fig:Fig1}
\end{figure}

\begin{figure}
\subfloat{
  \includegraphics[clip,height = 0.235\textwidth,width=0.235\textwidth]{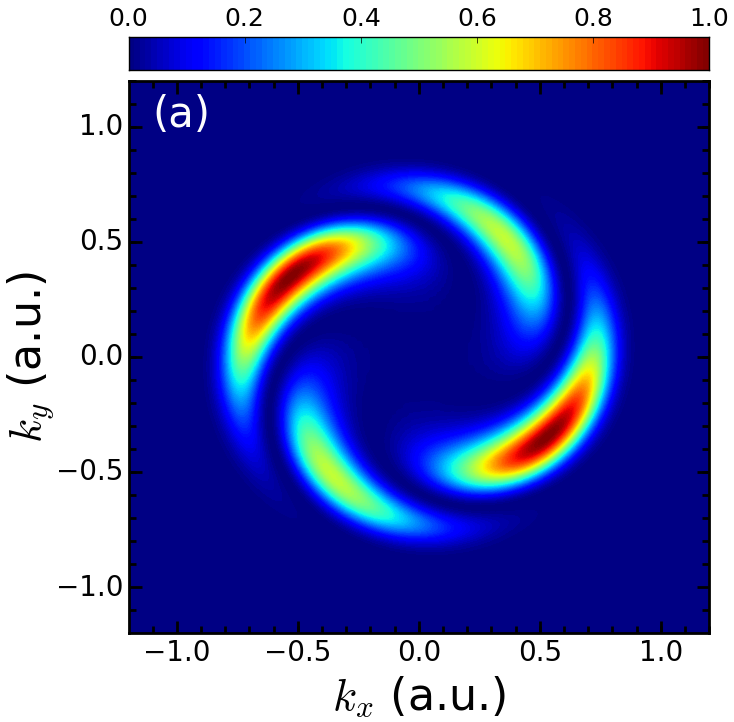}
}
\subfloat{
  \includegraphics[clip,height = 0.235\textwidth,width=0.235\textwidth]{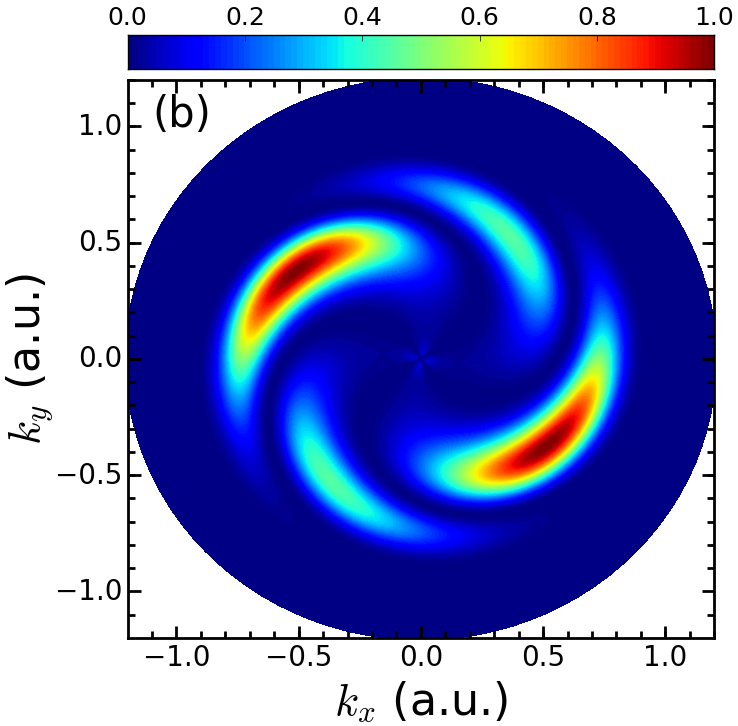}
}

\subfloat{
  \includegraphics[clip,height = 0.235\textwidth,width=0.235\textwidth]{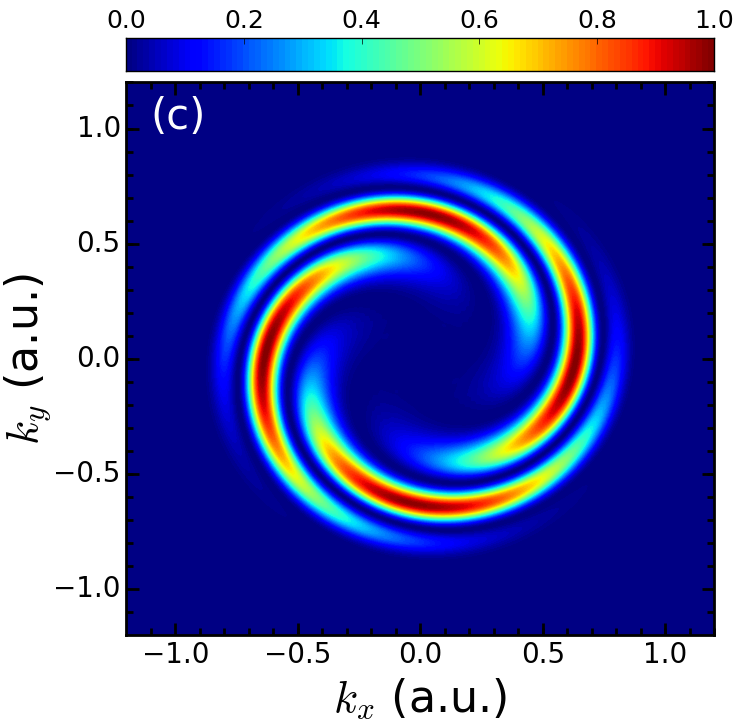}
}
\subfloat{
  \includegraphics[clip,height = 0.235\textwidth,width=0.235\textwidth]{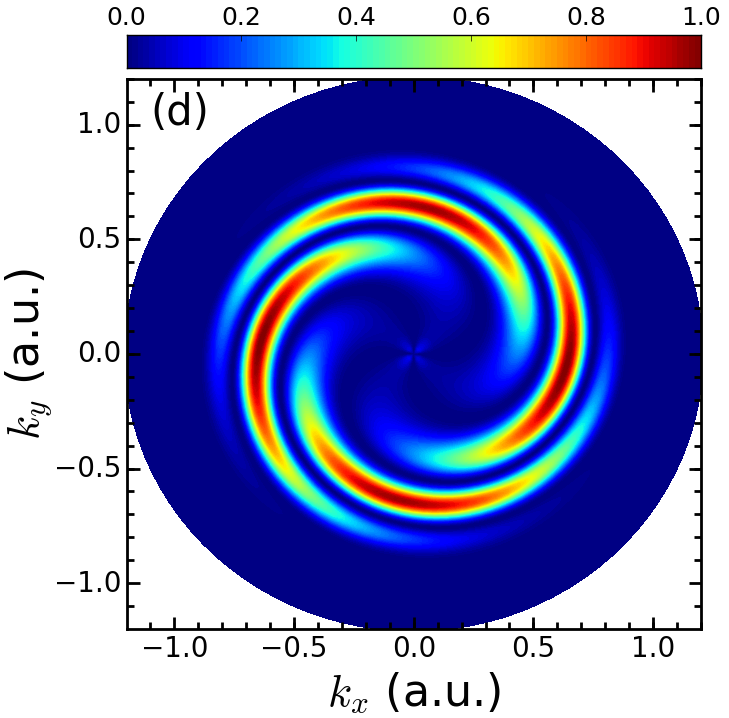}
}

\subfloat{
  \includegraphics[clip,height = 0.235\textwidth,width=0.235\textwidth]{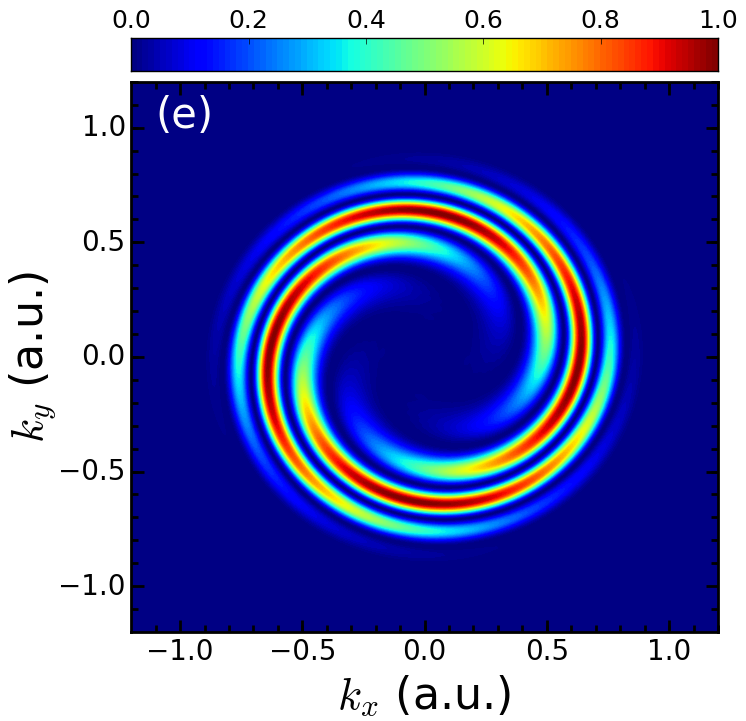}
}
\subfloat{
  \includegraphics[clip,height = 0.235\textwidth,width=0.235\textwidth]{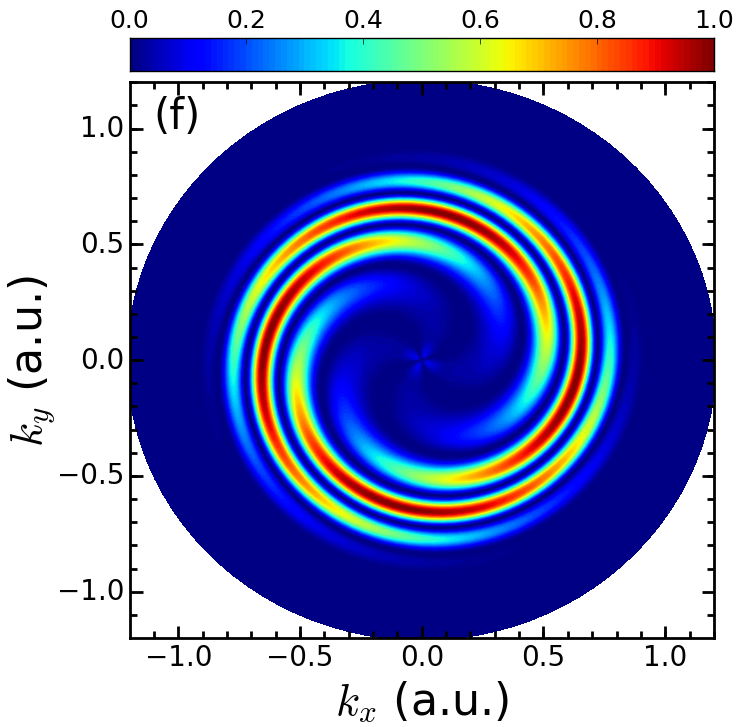}
}
\caption{(Color online) Photoelectron momentum distributions, in the polarization plane, following ionization of He by right- and left-hand circularly polarized laser pulses with non-zero time-delay $(\tau \neq 0)$.\ The distributions acquired through the generalized RMT method (for arbitrary polarization) are shown in (a), (c) and (e), whilst those obtained by Ngoko Djiokap {\it et al.}\ \cite{Djiokapetal2016}, employing a TDCC approach, are shown in (b), (d) and (f).\ Each pulse has a carrier frequency of $15\textrm{ eV}$, a duration of 6 cycles, a peak intensity of $1\times 10^{12}\,\textrm{Wcm}^{-2}$ and zero CEP $(\varphi_{1} = \varphi_{2} = 0)$.\ The pulses are delayed in time by $\tau = 500\textrm{ as}$ in (a) and (b), $\tau = 1.1\textrm{ fs}$ in (c) and (d), and $\tau = 1.65\textrm{ fs}$ in (e) and (f).\  Relative magnitudes are indicated by the color scales.}
\label{fig:Fig2}
\end{figure}

It should be highlighted that the momentum distributions, in each of Figures {\ref{fig:Fig1}} and {\ref{fig:Fig2}}, are all normalized in the same manner, such that the color scales extend from zero to one.\ This is due to a discrepancy in the ionization yields suggested by the present RMT approach and the TDCC calculations of Ngoko Djiokap {\it et al.}\ \cite{Djiokapetal2016}, the origins of which remain unclear.\ We note that the TDCC code developed by the latter authors, for two-electron systems in circularly polarized laser fields, is a generalization of a code employed in Ref.\ \cite{Djiokapetal2012} for pure linear polarization.\ Numerical results presented therein are well reproduced by the TDCC code used in Ref.\ \cite{Djiokapetal2016} for pulses with small ellipticity.\ Regarding the RMT data, our predicted yields have been verified in the specific case of Figure {\ref{fig:Fig1}}(a), where the laser field is linearly polarized in the $x$-direction (see below).\ For He, we expect to recover the same total yield irrespective of the orientation of the linear polarization axis, given the closed-shell nature of the system.\ In particular, for an analogous field distribution, but linearly polarized in the $z$-direction, calculations relying on the original RMT approach \cite{Mooreetal2011} reproduce the same yield to around $10^{-6}$.\ Moreover, for the photon energy $(15\textrm{ eV})$ in question, our calculated yields agree well (within $10\%$) with those estimated using established data for the generalized two-photon cross-section of He, acquired by means of previous $R$-matrix-Floquet calculations \cite{vdHBingham2005} and alternative {\it ab initio} methods \cite{Palaciosetal2008}.

To focus the present treatment purely on a comparison of the key qualitative features and their relative magnitudes, we normalize the RMT and TDCC spectra in a consistent fashion.\ By contrast, in Section {\ref{sec:He}}, we present energy and momentum distributions with a scale that reflects the true wavefunction density in momentum space.

We begin by discussing the results for zero time-delay $(\tau = 0)$ and two values of the relative CEP ($\varphi_{1} = 0, \varphi_{2} = 0$ and $\varphi_{1} = 0, \varphi_{2} = \pi/2$) in Eq.\ (\ref{TwopulsesHe}).\ Under these conditions, the superposition of two, counter-rotating, circularly polarized pulses yields a linearly polarized one, where the orientation of the polarization axis is determined by the relative CEP of the two pulses.\ In both Figures  {\ref{fig:Fig1}}(a) and (c), we find the expected quadrupole distribution of the ionized-electron momenta.\ For zero relative CEP, the field is linearly polarized along the $x$-axis, so that in Figure  {\ref{fig:Fig1}}(a), the photoelectron peaks are aligned along the $k_{x}$-axis, and the perpendicular $k_{y}$-axis.\ When the relative CEP is $\pi/2$, the polarization axis is rotated clockwise by $\pi/4$, giving rise to a disposition of the peaks shown in Figure {\ref{fig:Fig1}}(c).\ These properties of the spectra are entirely in line with the results of Ngoko Djiokap {\it et al.}\ \cite{Djiokapetal2016}.\ Irrespective of the CEP values, we find that the peak positions are consistent with the excess energy in a two-photon ionization event ${(k = {[2\left\{2\omega - E_{b}(\textrm{He})\right\}]^{1/2}} \approx 0.63\textrm{ a.u.})}$.\ Indeed, as emphasized in Ref.\ \cite{Djiokapetal2016}, the choice of pulse parameters (frequency $15\textrm{ eV}$, bandwidth $3.6\textrm{ eV}$ and peak intensity $1\times 10^{12}\,\textrm{Wcm}^{-2}$) ensure that two-photon ionization is, effectively, the only active transition pathway, so that the peak structures in Figure {\ref{fig:Fig1}} are uniquely attributable to this process.

Aside from the gross symmetry properties and peak dispositions, we also consider the relative intensities of the on- and off-axis emission features in Figure {\ref{fig:Fig1}}.\ In qualitative agreement with Ngoko Djiokap {\it et al.}\ \cite{Djiokapetal2016}, we find that the photoelectron peaks, aligned along the polarization axis of the field, are considerably more pronounced than those oriented perpendicular to it.\ Once more, this is an expected characteristic, arising due to interferences, among different ionization pathways, involving absorption of two photons with equal or opposite helicities.\ For on-axis emission, this interference is constructive, but becomes destructive for emission in the perpendicular direction, thus reducing the brightness of the off-axis features in Figure {\ref{fig:Fig1}}.\ More quantitatively, however, a notable difference appears in their relative magnitudes between the RMT and TDCC spectra.\ We predict the off-axis photoelectron peaks to be around three times smaller than those on-axis, whereas in the data of Ngoko Djiokap {\it et al.}\ \cite{Djiokapetal2016}, the peak heights differ by almost a factor of five.\ We have verified that this disparity does not arise from a deficiency in our description of the He target.\ Indeed, calculations employing more elaborate He$^{+}$ basis sets (e.g., incorporating the $\overline{3s}$, $\overline{3p}$ and $\overline{3d}$ pseudo-orbitals \cite{Hutchinsonetal2010,BrownvanderHart2012,ReyvdH2014}) yield no significant differences in the predicted momentum distributions.\ Moreover, for a field analogous to that of Figure {\ref{fig:Fig1}}(a), but linearly polarized in the $z$-direction, calculations exploiting the established RMT codes \cite{Mooreetal2011} suggest an almost identical spectrum in the $k_{x}k_{z}$-plane (with the same difference of peak heights as in Figures {\ref{fig:Fig1}}(a) and (c)).\ Of course, the same photoelectron emission properties should be recovered, for any orientation of the polarization axis, in the case of He, or any other closed-shell system.\ Notwithstanding this discrepancy in the RMT and TDCC predictions, the favourable comparison of the momentum spectra, for zero time-delay and both CEP values, highlights the reliability of our generalized RMT approach, at least for problems in which the axis of linear polarization is inequivalent to that of angular momentum quantization.

As discussed by Ngoko Djiokap {\it et al.}\ \cite{Djiokapetal2016}, non-zero time-delays, between the two counter-rotating pulses, ellicit the formation of multistart spiral vortex patterns in the photoelectron momentum distribution, whose characteristics depend sensitively on the relative handedness, phase and time-delay between the pulses.\ Such conditions thereby offer a more stringent test of accuracy for the present RMT methodology.

Figure {\ref{fig:Fig2}} compares the momentum distributions, computed by means of the present RMT method and the TDCC approach of Ngoko Djiokap {\it et al.}\ \cite{Djiokapetal2016}, for non-zero time-delays between the right- and left-hand circularly polarized pulses.\ In agreement with the TDCC data, we find four-start spiral patterns, with a counter-clockwise handedness.\ Moreover, with increasing time-delay between the pulses, we observe the same evolution in the number, and locations, of the maxima and minima in the distributions.\ Note that, for the shortest non-zero time-delay of $500\textrm{ as}$ (Figures {\ref{fig:Fig2}}(a) and (b)), the disparity in relative peak heights, found for the RMT and TDCC spectra in Figure {\ref{fig:Fig1}}, persists, but once more appears to constitute the only qualitative difference in the results.\ For longer time-delays, and in further concurrence with the data of Ngoko Djiokap {\it et al.}\ \cite{Djiokapetal2016} shown in Figures {\ref{fig:Fig2}}(d) and (f), the photoelectron peaks in Figures {\ref{fig:Fig2}}(c) and (e) assume a more complete four-fold rotational symmetry.\ In fact, the variation across the four maxima, in each of Figures {\ref{fig:Fig2}}(e) and (f), is less than $1\%$.\ Under these conditions, the RMT data of Figures {\ref{fig:Fig2}}(c) and (e), and the TDCC results in Figures {\ref{fig:Fig2}}(d) and (f), describe highly comparable photoelectron emission characteristics.

\section{Application to Single-photon Detachment from F$^{-}$ in a Circularly Polarized Light Field}
\label{sec:Fminus}

In Section {\ref{sec:He}}, we validated our new RMT methodology, for arbitrarily polarized laser fields, in the context of two-photon ionization of He by circularly polarized pulses.\ We now discuss a further application of the method, to the problem of single-photon detachment, from F$^{-}$, in a circularly polarized laser field.\

Our motivation for investigating this process is twofold.\ Firstly, F$^{-}$ constitutes an ionic and truly multielectron system, so that our treatment of the photodetachment dynamics therein emphasizes the generality of our approach with respect to the choice of target.\ Moreover, the theoretical description of negative ions presents an interesting challenge:\ both their structure, and field-driven dynamics, are influenced profoundly by multielectron correlations, particularly the strong dielectronic repulsion.\ A number of approximate methods are available to model photodetachment from complex negative ions \cite{GribakinKuchiev1997,Beiseretal2004,ShearerMonteith2013}, but tend to be rather limited in their account of electron repulsion.\ The role of multielectron correlations in negative-ion photodetachment was highlighted in a recent RMT study \cite{Hassounehetal2015}, addressing above-threshold detachment and strong-field rescattering in F$^{-}$.\ In the present work, the capacity of RMT to capture both long-range Coulomb interactions, as well as short-range exchange and correlation effects, is combined with a newly developed capability to treat atomic ionization dynamics in light fields of arbitrary polarization.\

Secondly, for fields of non-zero helicity, the dependence of the ionization characteristics on the atomic orbital phase, or sign of the single-electron magnetic quantum number $m_{l}$, has garnered substantial interest.\ On the one hand, it has long been recognized that circularly polarized fields preferentially ionize co-rotating electrons (i.e., positive $m_{l}$ for right-hand circular polarization), in both one-photon ionization and field-ionization from Rydberg states \cite{RzazewskiPiraux1993,Zakrzewskietal1993}.\ On the other hand, more recent experimental and theoretical works \cite{Herathetal2012,BarthSmirnova2011,BarthSmirnova2013,Lietal2015,Eckartetal2018} suggest that in the regime of strong-field tunnelling, non-adiabatic effects alter the characteristics of this $m_{l}$-dependence, whereby counter-rotating electrons (i.e., negative $m_{l}$ for right-hand circular polarization) are preferentially removed.\ Naturally, a better understanding of the origins of this behaviour, together with a systematic assessment of how, and when, a transition from one $m_{l}$-dependence to another emerges, constitute a matter of both fundamental and practical importance alike.\ We aim to address these questions in a future publication.\ Here, we report on the first step in such an investigation, employing our new RMT method to quantify the degree of quantum-state selectivity in the process of single-photon detachment, from F$^{-}$, in the field of a right-hand circularly polarized, femtosecond laser pulse.\

\subsection{Calculation Parameters}
\label{subsec:paramF}

Our treatment of the F$^{-}$ target in this work is as described in a previous RMT study \cite{Hassounehetal2015}, and is based upon that of much earlier $R$-matrix investigations of multiphoton detachment in this system \cite{vdH1996,vdH2000}.\ Within the inner region, we regard the ion as F to which is added a single electron.\ To describe the neutral F atom, we employ a set of Hartree-Fock $1s$, $2s$ and $2p$ orbitals, acquired for the F ground state from the data of Clementi and Roetti \cite{ClementiRoetti}, in conjunction with additional $\overline{3s}$, $\overline{3p}$  and $\overline{3d}$ pseudo-orbitals \cite{NiangDourneuf1997}.\ Inclusion of these pseudo-orbitals facilitates a more accurate determination of the $1s^{2}2s^{2}2p^{5}$ $^{2}P^{o}$ ground-state wavefunction, obtained in the form of a configuration-interaction expansion comprising the $1s^{2}2s^{2}2p^{5}$, $1s^{2}2s2p^{5}3s$, $1s^{2}2s^{2}2p^{4}3p$, $1s^{2}2s^{2}2p^{3}3p^{2}$ and $1s^{2}2s^{2}2p^{3}3d^{2}$ configurations.\ The present model suggests a binding energy of $E_{b}(\textrm{F}^{-})\approx 3.420\textrm{ eV}$ for the initial, $^{1}S^{e}$ F$^{-}$ ground state, which is similar to the experimental value of $3.401\textrm{ eV}$ \cite{Blondeletal2001}.

The radial extent of the inner-region is $50\textrm{ a.u.}$, which suffices to effectively confine the orbitals of the F$^{-}$ ion.\ The inner-region continuum functions are generated using a set of 60 $B$-splines of order 13 for each available orbital angular momentum of the outgoing electron.\ We retain all admissible $1s^{2}2s^{2}2p^{5} \epsilon l$ channels up to a maximum total orbital angular momentum $L_{\textrm{max}} = 9$, as well as all permitted magnetic substates with ${-9 \leq M_{L} \leq 9}$.\ The outer-region boundary radius ($3500\textrm{ a.u.}$), finite-difference grid spacing ($0.08\textrm{ a.u.}$) and time-step for the Arnoldi propagator ($0.01\textrm{ a.u.}$) are the same as those adopted in our calculations for He, as reported in Section {\ref{sec:He}}.

To probe the differential ionization dynamics of the $2p_{\pm 1}$ electrons in single-photon detachment, we subject the F$^{-}$ ion to a single, right-hand circularly polarized laser pulse, whose electric field is given by Eq.\ (\ref{field}) with ${\bf e} = ( \hat{\bf x} + i\hat{\bf y} )/\sqrt{2}$.\ The pulse is assumed to have a ramp-on/off temporal profile of the sine-squared form (\ref{sinesquared}), with a peak intensity $I_{0} = 1\times 10^{13}\,\textrm{Wcm}^{-2}$, a carrier frequency $\omega = 8\textrm{ eV}$, a CEP $\varphi = 0$ and a duration of $N=6$ cycles.

The photoelectron momentum distribution is computed via the method discussed in Section {\ref{subsec:paramHe}}.\ The distribution thereby obtained incorporates the emission characteristics of initially bound $2p$ electrons with ${m_{l} = -1,0}$ and 1, which we designate henceforth as $2p_{-1}$, $2p_{0}$ and $2p_{1}$ respectively.\ To decompose this total spectrum into its constituent $m_{l}$-selective components, our numerical projections should include only specific electron-detachment channels, identified by means of the following simple consideration.\ In a right-hand $(\eta = 1)$ circularly polarized laser field, the selection rule on the single-electron $m_{l}$ value is $\Delta m_{l} = 1$.\ Thus, in a single-photon detachment event, only those channels $p$, in which a final value of $m_{l_{p}} = 2$ can be realized, contribute to the spectrum for $2p_{1}$ electrons.\ Similarly, only those channels admitting a final value $m_{l_{p}} = 0$ (or $m_{l_{p}} = 1$) contribute to the spectrum for $2p_{-1}$ (or $2p_{0}$) electrons.\ Of course, this procedure is readily extended to multiphoton detachment processes.

\subsection{Results}
\label{subsec:F-Results}

Figure {\ref{fig:Fig3}} presents photoelectron momentum and energy distributions, in the polarization plane, for single-photon detachment from F$^{-}$, as driven by an $8\textrm{ eV}$, 6-cycle, right-hand circularly polarized laser pulse.\ As expected, the total distribution of Figure {\ref{fig:Fig3}}(a) exhibits a high degree of circular symmetry, and comprises a single, dominant ring of radius determined by the excess energy in this one-photon process (approximately $0.58\textrm{ a.u.}$).\ Note that the pulse peak intensity is too low to elicit higher-order (multiphoton) detachment with any substantial probability, so that the spectrum of Figure {\ref{fig:Fig3}}(a) displays primarily the single-photon feature.\ Nonetheless, a very faint outer ring is discernible, and is attributable to weak two-photon detachment (approximately $0.96\textrm{ a.u.}$).

\begin{figure}[h!]
\subfloat{
\includegraphics[clip,height = 0.44\textwidth,width=0.448\textwidth]{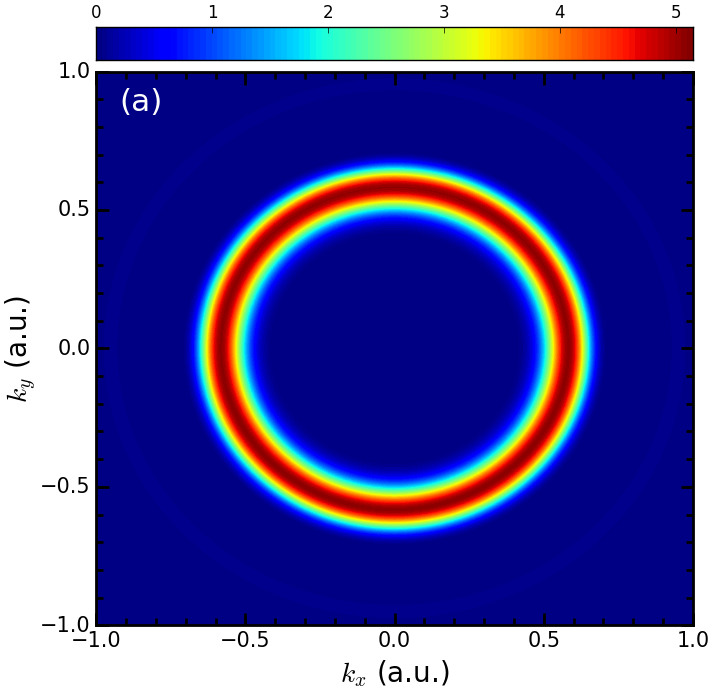}
}

\subfloat{
\includegraphics[clip,height = 0.39\textwidth,width=0.437\textwidth]{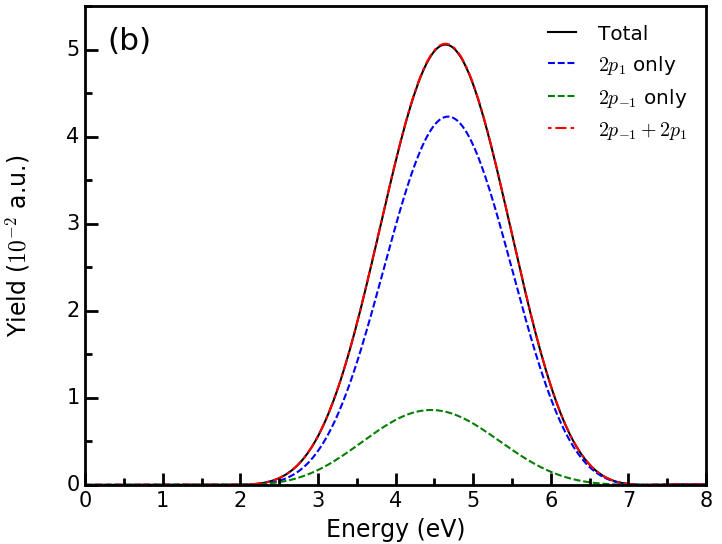}
}
\caption{(Color online) Photoelectron momentum and energy distributions, in the polarization plane, following single-photon detachment from F$^{-}$, initiated by a right-hand circularly polarized laser pulse with a carrier frequency of $\omega = 8\textrm{ eV}$, a duration of $N=6$ cycles, a peak intensity of $I_{0} = 1\times 10^{13}\,\textrm{Wcm}^{-2}$ and CEP $\varphi = 0$.\ (a) Total momentum distribution.\ Magnitudes are indicated by the color scale, and expressed in units of $10^{-2}\textrm{ a.u.}$.\ (b) Energy spectra for electrons ejected with zero azimuthal angle.\ The total spectrum (solid black line) is decomposed into individual spectra for the $2p_{1}$ (upper, dashed blue line) and $2p_{-1}$ (lower, dashed green line) electrons.\ The totality of the $2p_{1}$ and $2p_{-1}$ spectra (dash-dotted red line) is also shown.}

\label{fig:Fig3}
\end{figure}

The momentum distribution of Figure {\ref{fig:Fig3}}(a) is determined by the response of the $2p_{-1}$, $2p_{0}$ and $2p_{1}$ electrons of F$^{-}$ to the laser field (note that only the $2p$ electron emission channels are retained in the present calculations).\ To assess the sensitivity of single-photon detachment to the sign of the bound-electron magnetic quantum number, we decompose the distribution in Figure {\ref{fig:Fig3}}(a) following the procedure outlined in Section {\ref{subsec:paramF}}.\ The resulting momentum spectra possess the same qualitative (symmetry) properties as the total spectrum, and so to emphasize their quantitative differences, we discuss only one-dimensional energy spectra.\ Figure {\ref{fig:Fig3}}(b) displays the energy distributions for the $2p_{\pm 1}$ electrons of F$^{-}$, appropriate to a fixed photoelectron emission direction in the polarization plane (corresponding to zero azimuthal angle).\ For reference, the total spectrum for this direction is also shown in Figure {\ref{fig:Fig3}}(b), and corresponds to a one-dimensional slice of the two-dimensional distribution of Figure {\ref{fig:Fig3}}(a).\ Of course, the circular symmetry of the momentum distribution ensures that we may choose any emission direction, in the polarization plane, to investigate the energy spectra arising from ionization of these electrons.\ We observe that the total energy spectrum is essentially determined by ionization of the $2p_{\pm 1}$ electrons, which implies that the contribution of the $2p_{0}$ electrons is negligible.\ This behaviour is to be expected, and reflects the spatial orientation of the $2p_{0}$ and $2p_{\pm 1}$ electrons of the target ion.\ The symmetry axes of the $2p_{\pm 1}$ orbitals lie in the polarization plane, and as such, they are depleted with the highest probability.\ In contrast, the $2p_{0}$ orbital is aligned perpendicular to the polarization plane (i.e., its symmetry plane), and participates much more weakly.\ Moreover, the yield of $2p_{1}$ (co-rotating) electrons is almost five times larger than that of $2p_{-1}$ (counter-rotating) electrons.\ This dominance of the co-rotating electrons, in a single-photon process, has also been observed in previous studies of atomic hydrogen in microwave fields \cite{RzazewskiPiraux1993,Zakrzewskietal1993}, but is here demonstrated for a truly multielectron target in the XUV range.\ Our results therefore suggest that this $m_{l}$-selectivity is likely a fundamental attribute of single-photon ionization in fields of non-zero helicity, persisting not only in different wavelength regimes, but even in spite of dynamical, multielectron correlations in a more complex system.

\section{Conclusions}
\label{sec:Conclusions}

We have introduced an {\it ab initio} and fully non-perturbative RMT theory for ultrafast atomic processes in arbitrary light fields.\ Our approach represents the very latest evolution in time-dependent $R$-matrix techniques, retaining the same capacity as its predecessors \cite{Lysaghtetal2009,Mooreetal2011} in treating detailed, multielectron exchange and correlation effects, whilst facilitating the description of atomic ionization dynamics in truly multidimensional light-field configurations.\ These include, in particular, the fields arising from elliptically (and especially circularly) polarized laser pulses, for which compact and efficient radiation sources have become increasingly widespread.\ As such, our predictive capabilities should prove valuable in exploring the interplay between quantum many-body physics, and strong-field dynamics, in realistic and polarization-controllable laser fields.

Laser pulses with non-zero ellipticity effect atomic transitions in which the total orbital magnetic quantum number $M_{L}$ is not conserved.\ We have discussed the necessary alterations to both the RMT formalism (inner- and outer-region analyses), as well as the associated computer codes, to relax the constraint of $M_{L}$-conservation assumed in previous $R$-matrix techniques, and thereby enable an explicit account of all possible laser-induced transitions among magnetic substates of the target.\ Whilst modifications to the outer-region computational scheme are rather simple (requiring implementation of the long-range potentials (\ref{WDsum}) to (\ref{WPexpression}), and no changes to the domain decomposition parallelisation strategy), substantial alterations to the inner-region scheme were essential.\ In particular, to facilitate the numerical solution of the system of equations (\ref{ODEs}), with the Hamiltonian (\ref{Hamarb}) appropriate for a field of arbitrary polarization, we have modified the inner-region parallelisation structure, and developed a much more robust set of communication routines for the efficient distribution, and exchange, of both Hamiltonian-matrix and wavefunction data.\ We emphasize that our strategy for such communications is now based solely on the $LM_{L}S\pi$ couplings (selection rules) relevant to the laser field polarization of interest, and assume no fixed structure of the Hamiltonian matrix.\ Our scheme could be extended to manage the communications required for other interactions, such as those of a non-dipole nature, and has already been adapted for time-dependent molecular $R$-matrix calculations.\ As such, the computational progress reported here is not only relevant to atomic RMT calculations for arbitrarily polarized light fields, but bears important implications for future evolutions and applications of the RMT methodology.

As a first demonstration of our generalized RMT approach, we investigate the formation of multistart, spiral vortex features in the photoelectron momentum distributions of He, irradiated by a pair of time-delayed, ultrashort, circularly polarized laser pulses with opposite helicities.\  Through comparison of the RMT data with the TDCC results of Ngoko Djiokap {\it et al.} \cite{Djiokapetal2016}, we have verified that our calculations can reproduce the key qualitative features of the photoelectron momentum distributions in the polarization plane, correctly capturing the sensitivity of the electron vortex properties (number and orientation of the spiral arms) to the relative handedness, CEP and time-delay of the pulses.\ Our predicted ionization yields, in cases where the superposition of the two circularly polarized pulses yields a linearly polarized one, are supported by available data for the generalized two-photon cross-section of He.

The predictive capacity of our latest RMT approach has been further underlined in a study of single-photon detachment from F$^{-}$, initiated by a single, right-hand circularly polarized, femtosecond laser pulse.\ We highlight that this application relies on both the intrinsic ability of RMT to capture many-body exchange and correlation effects, as well as our newly developed capability to treat atomic ionization in light fields of arbitrary polarization.\ To assess the sensitivity of the photodetachment dynamics to the sign of the bound-electron magnetic quantum number $m_{l}$, we have decomposed the photoelectron energy spectrum into its $m_{l}$-selective components.\ Our results suggest that the ionization response of co-rotating ($2p_{1}$) electrons dominates that of counter-rotating ($2p_{-1}$) electrons.\ Such behaviour was previously identified in studies of atomic hydrogen exposed to microwave fields \cite{RzazewskiPiraux1993,Zakrzewskietal1993}, but has here been evidenced for a truly multielectron target in the XUV range.\ The latter observation may suggest that preferential removal of electrons, with one sign of $m_{l}$, is a fundamental attribute of single-photon ionization in fields with non-zero helicity, persisting not only in different wavelength regimes, but even in spite of dynamical, multielectron correlations in more complex systems.

More generally, the dependence of the ionization characteristics on the atomic orbital phase, or sign of $m_{l}$, has been the subject of substantial research activity.\ Recent experimental and theoretical works \cite{Herathetal2012,BarthSmirnova2011,BarthSmirnova2013,Lietal2015,Eckartetal2018} suggest that in the regime of strong-field tunnelling, non-adiabatic effects alter the $m_{l}$-dependence observed in this work, whereby counter-rotating electrons (i.e., negative $m_{l}$ for right-hand circular polarization) are preferentially ionized.\ The generalized RMT approach, introduced in this article, represents a viable theoretical tool for investigating this transition, in a systematic fashion, as the driving wavelength (or number of photons required for ionization) increases.\ However, we emphasize that the newly developed suite of codes appear promising for a plethora of other novel applications, whether in regard of fundamental, laser-induced atomic processes (in particular, inner-shell dynamics \cite{Snelletal1996} and the production of valence ring currents \cite{Eckartetal2018}), or experimental schemes of contemporary interest (including the attoclock \cite{Eckleetal2008,Eckleetal2008Science,Staudteetal2009,Pfeifferetal2012,Wuetal2012,Torlinaetal2015}, HHG in cross-polarized \cite{Lambertetal2015,Soiferetal2013} and circularly or elliptically polarized \cite{Milosevicetal2000,Fleischeretal2014,Medisauskasetal2015,Kfiretal2015} laser pulses, as well as attosecond photoelectron holography \cite{Poratetal2018}).\ As a result, the methodology presented here constitutes a significant and timely development in $R$-matrix techniques, facilitating the accurate simulation, and more profound understanding, of ultrafast, many-body dynamics in atomic systems exposed to arbitrarily polarized light fields.

The data presented in this article may be accessed using Ref.\ \cite{PUREdatasets}.\ The RMT code is part of the UK-AMOR suite, and can be obtained for free through Ref.\ \cite{RMTrepo}.

\begin{acknowledgements}

We wish to thank Prof. J. M. Ngoko Djiokap for provision of the TDCC data presented in Figures {\ref{fig:Fig1}} and {\ref{fig:Fig2}}, as well as for insightful discussion surrounding the comparison with RMT results.\ We also express our gratitude to Drs.\ J. Benda and Z. Ma{\v{s}}{\'{i}}n for their collaboration in developing and maintaining the code found using Ref.\ \cite{RMTrepo}.\ This work benefited from computational support by CoSeC, the Computational Science Centre for Research Communities, through CCPQ.\ DDAC acknowledges financial support from the UK Engineering and Physical Sciences Research Council (EPSRC).\ ACB, HWvdH and GSJA acknowledge funding from the EPSRC under grants EP/P022146/1, EP/P013953/1 and EP/R029342/1.\ The calculations reported in this work relied on the ARCHER UK National Supercomputing Service (\href{http://www.archer.ac.uk/}{\texttt{www.archer.ac.uk}}).

\end{acknowledgements}

\end{document}